\documentclass[fleqn,usenatbib]{mnras}

\usepackage[T1]{fontenc}

\DeclareRobustCommand{\VAN}[3]{#2}
\let\VANthebibliography\thebibliography
\def\thebibliography{\DeclareRobustCommand{\VAN}[3]{##3}\VANthebibliography}

\usepackage{graphicx}	% Including figure files
\usepackage{amsmath}
\usepackage{amssymb}	% Extra maths symbols
\usepackage{xcolor}
\usepackage{hyperref}
\usepackage{gensymb}
\usepackage{xspace}%%%%%%%%%%%%%%%%%%%%%%%%%%%

\usepackage{newtxtext,newtxmath}
\newcommand{\Dskel}{D$_{\mathrm{skel}}$}
\newcommand{\mstar}{M$_{\star}$}
\newcommand{\Msun}{M$_{\odot}$}

\newcommand{\Mbar}{M$_{\mathrm{bar}}$ / M$_{\odot}$}
\newcommand{\Mhalo}{M$_{\mathrm{halo}}$}

\newcommand{\mhi}{\ensuremath{\text{M}_{\rm HI}}\xspace}

\newcommand{\hi}{\textsc{H$\,$i}\xspace}
\newcommand{\kms}{\ensuremath{\text{km}\,\text{s}^{-1}}\xspace}

\title[The Cosmic Web in RESOLVE and ECO]{The effect of cosmic web filaments on galaxy properties in the RESOLVE and ECO surveys}

\author[M. Hoosain et al.]{
Munira Hoosain,$^{1,2}$\thanks{E-mail: munira@saao.ac.za}
Sarah-L. Blyth,$^{1}$
Rosalind E. Skelton$^{2}$, Sheila J. Kannappan$^{3}$, \newauthor
David V. Stark$^{4}$, Kathleen D. Eckert$^{3}$, Zackary L. Hutchens$^{3}$, Derrick S. Carr $^{3}$ and \newauthor Katarina Kraljic$^{5}$ 
\\
$^{1}$Department of Astronomy, University of Cape Town, Private Bag X3, Rondebosch,  7701, Cape Town, Republic of South Africa \\
$^{2}$South African Astronomical Observatory, 1 Observatory Road, Cape Town, 7925, Republic of South Africa \\
$^{3}$ Department of Physics \& Astronomy, CB3255, University of North Carolina, Chapel Hill, 27599, USA \\
$^{4}$ Space Telescope Science Institute, 3700 San Martin Dr Baltimore, MD, 21218, USA \\
$^{5}$ Université de Strasbourg, CNRS UMR 7550, Observatoire astronomique de Strasbourg, 11 rue de l’Université, 67000 Strasbourg, France \\}

\date{Accepted XXX. Received YYY; in original form ZZZ}

\pubyear{2022}

\begin{document}
\label{firstpage}
\pagerange{\pageref{firstpage}--\pageref{lastpage}}
\maketitle

\begin{abstract}
Galaxy environment plays an important role in driving the transformation of galaxies from blue and star-forming to red and quenched. Recent works have focused on the role of cosmic web filaments in galaxy evolution and have suggested that stellar mass segregation, quenching of star formation and gas-stripping may occur within filaments. We study the relationship between distance to filament and the stellar mass, colour and \hi gas content of galaxies using data from the REsolved Spectroscopy of a Local VolumE (RESOLVE) survey and Environmental COntext (ECO) catalogue, two overlapping census-style, volume-complete surveys. We use the Discrete Persistence Structures Extractor (DisPerSE) to identify cosmic web filaments over the full ECO area. We find that galaxies close to filaments have higher stellar masses, in agreement with previous results. Controlling for stellar mass, we find that galaxies also have redder colours and are more gas poor closer to filaments. When accounting for group membership and halo mass, we find that these trends in colour and gas content are dominated by the increasing prevalence of galaxy group environments close to filaments, particularly for high halo mass and low stellar mass galaxies. Filaments have an additional small effect on the gas content of galaxies in low-mass haloes, possibly due to cosmic web stripping.

\end{abstract}

\begin{keywords}
galaxies: evolution -- galaxies: groups: general -- (cosmology:) large-scale structure of Universe
\end{keywords}

%%%%%%%%%%%%%%%%%%%%%%%%%%%%%%%%%%%%%%%%%%%%%%%%%%

%%%%%%%%%%%%%%%%% BODY OF PAPER %%%%%%%%%%%%%%%%%%

\section{Introduction} \label{sect:intro}

Early wide-field spectroscopic galaxy surveys \citep[e.g.][]{Colless20012dF, Jones20046dF, Eisenstein2011AJSDSS} revealed that galaxies form an intricate, large-scale network of massive superclusters connected by filamentary structures, surrounding near-empty voids, known as the Cosmic Web \citep[e.g.][]{deLapparentCosmicWeb1986, bond1996filaments}. In the standard model of cosmology, the large scale structure observed today is a consequence of two mechanisms \citep{bond1996filaments}: fluctuations in the initial matter density field shortly after the Big Bang, and hierarchical structure formation. 

The environment in which galaxies are located affects their evolution. Elliptical and S0 galaxies are more abundant in dense regions such as clusters, while spiral and irregular galaxies dominate the population in low density regions \citep[morphology-density relation][]{dressler1980morphologydensity}. In dense cluster environments, various processes impact the star formation properties and morphologies of galaxies. For example, ram-pressure stripping \citep{GunnGott1972}, where cold gas from the disk is removed as a galaxy falls through the hot, dense intra-cluster medium (ICM), harassment, either through tidal interactions with neighbouring galaxies or the cluster itself, and galaxy mergers can result in quenching of star formation as well as morphological disruptions. Strangulation \citep{Balogh2000Strangulation1, Balogh2000Strangulation2}, where the extended halo gas reservoir is removed as a galaxy falls into a cluster, also leads to redder, more passive cluster populations compared to bluer, star forming field galaxies. 

While observational studies have typically concentrated on galaxy clusters and groups as environments that drive galaxy evolution, recent work has suggested that galaxy properties can be affected due to their location inside, or close to, cosmic web filaments. \citet{Donnan2022metal} described two pathways for the cosmic web to influence the properties of galaxies: by affecting the growth of their haloes, or by affecting the gas content of galaxies.  

One of the most consistent trends within the literature is that the stellar mass of galaxies in filaments is typically higher than galaxies outside of filaments \citep{Laigle2018cosmos, Chen2017filaments, Luber2019CHILES, malavasi2017vimos}. \citet{Kuutma2017filaments} found an increase in the fraction of early-type galaxies close to filaments. Additionally, \citet{santiago2020identification} found an increase in the elliptical to spiral ratio close to filaments, indicating that filaments - as intermediate density regions - follow the morphology-density relation.

Previous studies found that galaxies were redder within filaments compared to galaxies outside of filaments \citep[e.g.][]{Kuutma2017filaments, Chen2017filaments, kraljic2018,Luber2019CHILES}. \citet{Kuutma2017filaments} and \citet{kraljic2018} found that these redder colours corresponded to a decrease in the specific star formation rates for galaxies in filaments. Additionally, increases in the fraction of passive (i.e. non-star forming) galaxies within filaments were found, indicative of quenching due to the filaments \citep{Sarron2019preprocessing, blue2020chiles, kraljic2018} . Evidence of  lower sSFR and higher stellar metallicity was also found for central galaxies in filaments, suggesting that filaments play a role in quenching \cite[e.g.][]{Winkel2021quenching, kraljic2018, Donnan2022metal}. 

While several studies have examined the stellar properties of galaxies in filaments, more observations are needed to understand the gas properties of filament galaxies. \citet{kleiner2016DisPerSE} showed that galaxies with $\log$ (\mstar / \Msun) $> 11$  had higher \hi  masses within filaments compared to galaxies in a control sample outside of filaments, when correcting for density, which they attributed to gas accretion from filaments. However, using data from the ALFALFA survey \citep{giovanelli2005ALFALFA}, \citet{odekon2018effect} found that in a lower mass regime ($8.5$ $<$ $\log$(\mstar/\Msun)~$< 10.5$), galaxies lose \hi  gas as they enter filaments and redden as they are quenched, seemingly in tension with the result by \citet{kleiner2016DisPerSE}.  In the SIMBA simulations \citep{dave2019simba}, \citet{Bulichi2023SIMBA} found that cold gas is suppressed close to filaments, but that the complexity of associating HI with particular galaxies in dense environments requires further investigation. Using preliminary observations from the CHILES survey and photometric \hi  estimations, \citet{Luber2019CHILES} and \citet{blue2020chiles} suggest that the gas fractions of galaxies may increase away from filaments. However, the differing mass ranges between the studies on the gas content of galaxies with respect to filaments may suggest that stellar mass affects whether galaxies replenish or lose gas from the cosmic web \citep{odekon2018effect}.

Observing the spin-alignment of galaxies with respect to nearby filaments is important for studies testing the Tidal Torque Theory \citep{hoyle1949TTT,Peebles1969TTT, Doroshkevich1970TTT, White1984TTT} and its role in gas accretion for galaxies \citep{Laigle2015TTT, Laigle2018cosmos}. The spin alignment (or mis-alignment) of galaxies is mass-dependent and may allow galaxies to accrete gas from the cosmic web, as shown by simulations \citep{kraljic2020spin, Laigle2015TTT, Laigle2018cosmos}. This alignment has been observed with varying results \citep[see][]{Tempel2013spinalign, TempelLibeskind2013, Welker2019spin, blue2020chiles, Barsanti2022SPIN}. This is important for tracing mergers and their subsequent effect on changing the angular momentum of galaxies. More recent work using a small sample of galaxies from the MIGHTEE-\hi \citep{jarvis2017MIGHTEE} survey found that galaxies with low gas fractions (log (\mhi/\mstar) < 0.11)  are more likely to have their spin aligned with cosmic web filaments than galaxies with higher gas fractions \citep{tudorache2022mighteespin}.  However, results from simulations suggest that low halo-mass galaxies may be subject to `cosmic web stripping', where the ram-pressure of the filament environment overcomes the lower binding energy of the haloes and removes gas from galaxies \citep{Benitez-Llambay2013cosmicwebstripping, Thompson2022strippingvoids}.

While the spin and gas properties are related, more observations are needed to understand the complex processes that affect the gas content and refuelling of galaxies with respect to their location in the cosmic web.  It is important to note that few of the studies of galaxy evolution in filaments distinguish between filaments and galaxy groups, which may be embedded in filaments and drive similar observational trends. Groups are known sites of star-formation quenching \citep{peng2010mass, cluver2020GAMAgroups}. This may lead to `pre-processing' of galaxies, where quenching, reddening and morphological changes occur in groups as they travel along filaments before entering the cluster environment \citep{fujita2004pre, Sarron2019preprocessing}. Further, \citet{Song2021filaments} found that the host halo of a galaxy is primarily responsible for driving trends in galaxy properties close to filaments, with secondary effects due to the filaments themselves. An outstanding question is: are the observed properties of galaxies in filaments a result of their group membership and halo properties, or due to the filament environment and its associated processes?

In this paper, using data from the REsolved Spectroscopy Of a Local Volume \citep[RESOLVE;][]{Eckert2015RESOLVEphoto,stark2016resolve, KannappanWei2008RESOLVE}  survey and Environmental COntext (ECO) \citep{Moffett2015ECO} catalogue, we identify cosmic web filaments in the local universe and investigate the properties of galaxies with respect to their proximity to filaments. The multi-wavelength data and value-added catalogues from RESOLVE and ECO enable us to study the distributions of stellar mass, colour, and \mhi/\mstar  gas fractions (G/S) of galaxies down to the ECO baryonic mass completeness limit. This dataset allows us to take into account the membership of galaxies to groups or clusters as we aim to disentangle the effects of the filament environment on galaxy evolution.

Throughout this work, the standard $\Lambda$CDM model is assumed with $\Omega_{\mathrm{M}}$ = 0.3, $\Omega_{\mathrm{\Lambda}}$ = 0.7 and H$_{0}$ = 70 km/s/Mpc. Section \ref{sect:data} provides an overview of the data used in this project from RESOLVE and ECO. Section \ref{sect:method} provides an overview of the process used to detect filaments. The analysis of galaxy properties with respect to filaments and the results are presented in Section \ref{sect:results}. Finally, Section \ref{sect:discussion} discusses the results in context and the summary, conclusion and outlook for future work are presented in Section \ref{sect:conclusion}.

\section{Data} \label{sect:data}

\subsection{RESOLVE and ECO}

\begin{figure*}
\includegraphics[width=2\columnwidth]{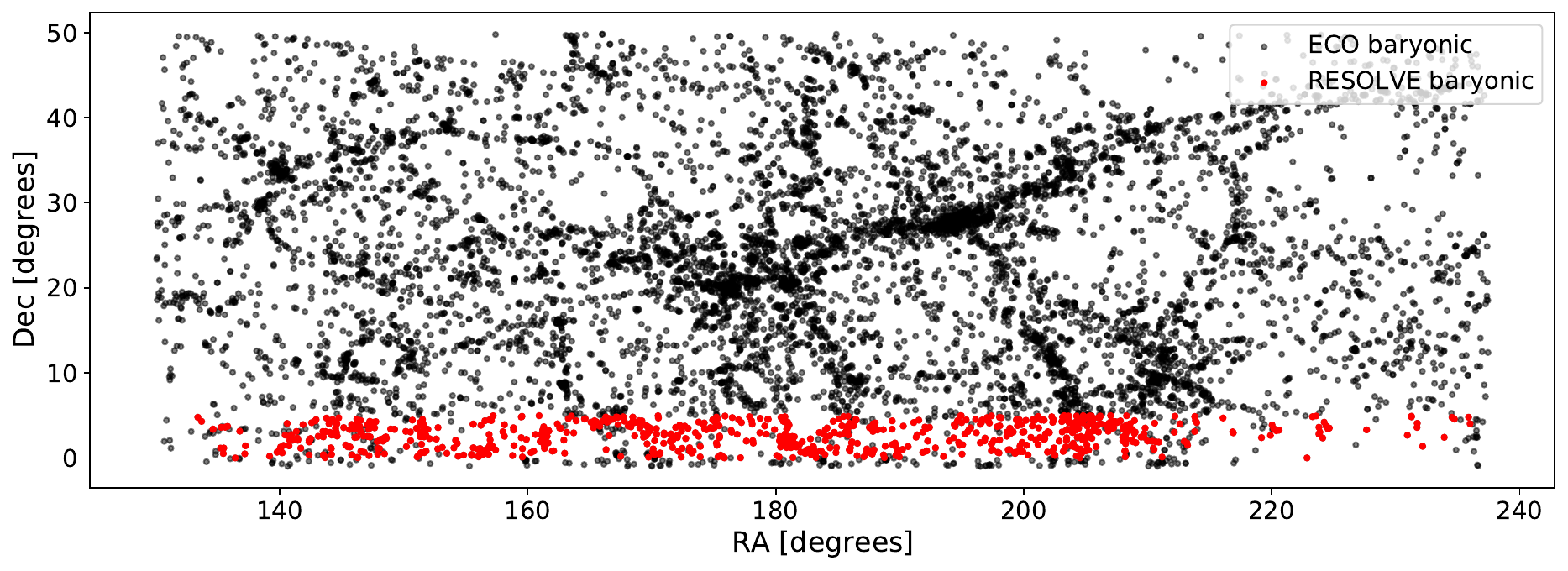}

\caption{The distribution of galaxies in the ECO survey (gray) in comparison to the RESOLVE-A survey (red). RESOLVE-A groups fall within 4500 \kms ~<~ $cz$ ~<~7000 \kms.. This figure only shows galaxies which belong to the baryonic mass selected sample. Note that ECO spans a larger range of redshifts than RESOLVE, with group velocities of 2530 \kms < $cz$ < 7470 \kms.}
\label{fig:resolveECO}
\end{figure*}

%rephrase this paragraph
The analysis of galaxy properties with respect to filaments in this paper uses observational data from the RESOLVE survey \citep{KannappanWei2008RESOLVE, Eckert2015RESOLVEphoto,stark2016resolve} and ECO catalogue \citep{Moffett2015ECO, eckert2016resolve} and with updates described in \citet{Hutchens2023G3}. RESOLVE is a volume-limited census of the local universe which is highly complete down to low mass, gas rich galaxies (M$_{\mathrm{bary}} \sim $ $10^{9.1} - 10^{9.3}$ \Msun), where M$_{\mathrm{bary}}$ = (\mstar + 1.4 M$_{\mathrm{HI}}$). RESOLVE-A spans 131.25\textdegree ~<~ R.A. ~<~  236.25\textdegree, 0\textdegree ~<~ Dec. ~<~ 5\textdegree, and 4500 \kms ~<~ $cz$ ~<~7000 \kms. ECO encompasses a larger field at 130.05\textdegree~<~R.A.~< 237.45\textdegree\ and -1\textdegree~<~Dec.~<~+49.85\textdegree, with galaxy velocities 4500 \kms < $cz$ < 7000 \kms, spanning a volume greater than 400~000 Mpc$^3$. ECO surrounds RESOLVE-A, which contains galaxies with velocities 2530  \kms < $cz$ < 7470 \kms, forming a 1 Mpc buffer around it. Figure \ref{fig:resolveECO} shows the sky distribution of galaxies in ECO and RESOLVE-A. Both ECO and RESOLVE are accompanied by comprehensive group catalogues~\citep{eckert2016resolve,Eckert2017ECOgroups} which allow the three-dimensional positions of galaxies to be corrected for redshift space distortions. In this study, we use ECO-DR3 \citep{Hutchens2023G3}, which contains the RESOLVE-A field and all data therein. 

\subsection{Spectroscopic redshifts}
The ECO DR3 catalogue update \citep{Hutchens2023G3} includes redshifts compiled from the Sloan Digital Sky Survey \citep[SDSS][]{York2000SDSS, alam2015SDSS}, HyperLEDA \citep{Paturel2003HyperLEDA}, GAMA \citep{Driver2011GAMA}, 2dF \citep{Colless20012dF}, 6dF \citep{Jones20046dF}, and RESOLVE's own observations. The spectroscopic redshifts are crucial for measuring accurate galaxy positions for 3D filament finding.

\subsection{Photometric data}
The data for ECO and RESOLVE were reprocessed using custom pipelines for consistency in the galaxy properties, as described by \citet{Moffett2015ECO} and \citet{eckert2016resolve}.The RESOLVE photometric data and derivation of galaxy properties are discussed in \citet{Eckert2015RESOLVEphoto}. Because RESOLVE-A lies within ECO, the RESOLVE galaxy properties are used for galaxies in the overlapping area.  We use galaxy \textit{u-r} colours, r-band magnitudes, stellar masses, halo masses and \hi gas masses for this analysis. The stellar masses and \textit{u-r} colours are derived from SED modelling and have external extinction and k-corrections applied. The modelling and thus the stellar mass incorporates internal extinction corrections, but we use colours only corrected for foreground extinction for consistency with other observational studies. 

The sample for this work is drawn from the ECO volume, including RESOLVE-A, and is baryonic-mass selected and complete to log (\Mbar)> 9.3. The choice of baryonic mass allows us to probe galaxies which may have low stellar masses but high gas-richness and follows previous work in RESOLVE such as \citet{stark2016resolve}. In total, the sample consists of $\sim$ 9612 galaxies.

\subsection{\hi data}

A deeper census of the gas content of galaxies in the RESOLVE survey volume of was conducted by \citet{stark2016resolve}, who compiled existing ALFALFA data \citep{giovanelli2005ALFALFA} and \hi measurements from other sources, as well as completed dedicated observing campaigns on the Robert C. Byrd Green Bank Telescope (GBT) and Arecibo Telescope. Value added data such as upper limit estimates and confusion flags were added by the RESOLVE and ECO teams \citep{Hutchens2023G3,stark2016resolve}. 
In this work, the ALFALFA \hi observations are from the ALFALFA-100 survey \citep{haynes2018alfalfa100} which was cross-matched with ECO galaxies as part of ECO DR3 (see \citet{Hutchens2023G3}). ECO measurements within the RESOLVE-A volume are substituted with their RESOLVE values from \citet{stark2016resolve} where available. Photometric gas fraction estimates are used where direct, high signal-to-noise (S/N > 5) detections were unavailable and for highly confused \hi measurements. Photometric gas fractions are calculated using an updated version of the tight correlation between observed galaxy colour and gas fraction \citep{Kannappan2004gastostars}. This technique uses the correlation between $u-J$ SED modelled colour and log (\mhi / \mstar) for galaxies with well constrained \hi\ measurements, taking into account the $b/a$ axis ratio, to predict the gas fraction using photometric parameters \citep{Eckert2015RESOLVEphoto}. This corresponds to the \texttt{logmgas} column within the ECO DR3 catalogue.

\subsection{Groups}

The galaxy groups used in this study are drawn from ECO DR3 \citep{Hutchens2023G3} using the group finding methodology originally described in \citet{Eckert2017ECOgroups}.

\citet{Hutchens2023G3} produced two group catalogues for the ECO survey: an updated Friends-of-Friends catalogue based on \citet{Eckert2017ECOgroups}, and a novel 'G3' group catalogue which identified groups by using giant galaxies as a basis (see section \ref{subsec:groupssingles} for a discussion of our choice of group finding algorithm). In this work, we use groups which are identified using the Friends-of-Friends technique \citep{berlind2006FoF} with a tangential linking length of 0.07 and line-of-sight linking length of 1.1 \citep{duarte2014FoF, Eckert2017ECOgroups}, followed by a procedure to split false FOF pairs. Halo masses are calculated using abundance matching techniques in \citet{Eckert2017ECOgroups} and the brightest galaxy in each group is designated as its central galaxy. 

\section{Method} \label{sect:method}

In this paper, we use the Discrete Persistence Structures Extractor (DisPerSE) \citep[DisPerSE;][]{sousbie2011disperseI,sousbie2011disperseII} to identify cosmic web filaments in the ECO field. DisPerSE can be applied to observational galaxy position data. However, the redshift or recessional velocity data may be subject to redshift space distortions \citep{deLapparentCosmicWeb1986}. This `Finger of God' effect is corrected by assigning the group velocity $cz_{\mathrm{group}}$ to each galaxy, which effectively collapses the stretched out `fingers'.

At low redshift, the distance \textit{R} $\approx$ $cz_{\mathrm{group}}$  $\mathrm{[km/s]}$ / H$_0 \mathrm{[km/s / Mpc]}$. Additionally, galaxy sky positions are converted to co-moving coordinates using the following transformations with units of megaparsecs (Mpc) from the observer: 

\begin{equation}
\begin{aligned}
    X = R \sin(90\degree -\rm{Dec}) \cos(\rm{RA}); \\
    Y = R \sin(90\degree -\rm{Dec}) \sin(\rm{RA}); \\
    Z = R \cos(90\degree - \rm{Dec}). \\
\end{aligned}
\label{Eq:comoving}
\end{equation}

\subsection{DisPerSE}

DisPerSE is a scale-free and parameter-free software that takes advantage of Discrete Morse Theory, which is a topological tool to map large scale structure features, and Persistence Theory to measure the robustness of identified features. DisPerSE uses the Delaunay Tesselation Field Estimator to construct the density field of a galaxy distribution. The software then applies Discrete Morse Theory to construct the Morse-Smale Complex and identifies changes in gradient in the density field. By using a persistence threshold as a measure of robustness, DisPerSE outputs the most robust large scale structure features such as filaments, voids, walls and nodes. Because DisPerSE has these advantages and is widely used in studies examining the effect of the cosmic web on galaxy properties, it is ideal for this study. 

We selected to use the \texttt{smooth} boundary condition when evaluating the density field using the Delaunay Tessellation Field Estimator in 3D since this is the recommended option for distributions with irregular shapes.
The density field is used to determine the Morse-Smale Complex and the filament `skeleton' is then extracted. We used a $5\sigma$ persistence threshold.

\subsection{Calculating Distance to Filaments}\label{sec:FilDist}

\begin{figure}
\includegraphics[width=\columnwidth]{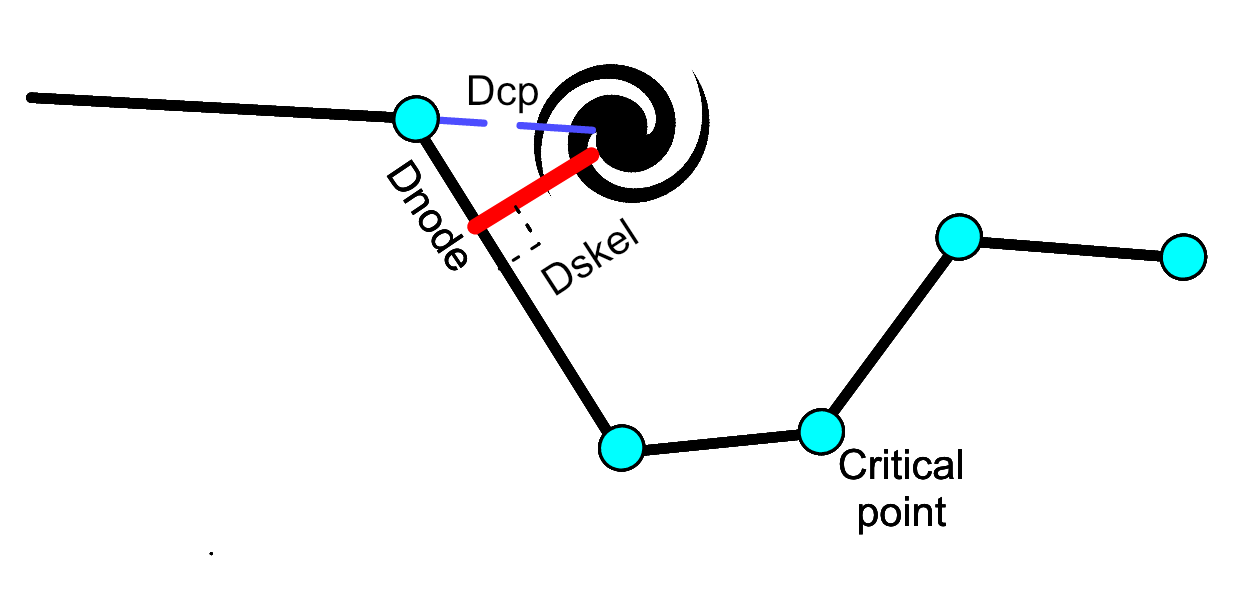}

\caption{Illustration of the difference between the D$_{\mathrm{cp}}$, parameter (dashed, blue line), which shows the distance to the nearest critical point, and the \Dskel (red, solid line) parameter, which measures the perpendicular distance to the nearest filament segment. Critical points at the end points of filament segments are marked with cyan circles and the filament segments are represented by the solid black lines. }
\label{fig:DcpDskeldiagram}
\end{figure}

When studying the effect of filaments on galaxy evolution, one useful parameter is the distance from a galaxy to the nearest filament. In previous studies using DisPerSE, this has been measured as the distance to the nearest critical point, D$_{\mathrm{cp}}$  \citep[for e.g.][]{Luber2019CHILES, blue2020chiles}, or, alternatively, as the perpendicular distance to the nearest filament segment or ‘skeleton’, (\Dskel )  \citep[see][]{kraljic2018}. These metrics are illustrated in the diagram in Figure \ref{fig:DcpDskeldiagram}.

In our analysis our preferred parameter is \Dskel\ but we also calculate D$_{\mathrm{cp}}$ for each galaxy to enable comparison with other studies.

\section{Results} \label{sect:results}

\subsection{The effect of filaments on galaxy properties}\label{sec:galprop}

To investigate the effects of the filament environment on galaxy properties, we examine trends in galaxy stellar mass, colour and \hi\ gas content as a function of distance from filaments. Knowing that cluster and group environments affect galaxy properties and that galaxy groups can be located along the filament backbones, we then attempt to distinguish between the effects caused by the filament and group environments by sub-dividing our sample into group and non-group galaxies and comparing the properties of the sub-samples in different stellar mass bins.

\begin{figure*}
\centering
    \includegraphics[width=0.8\linewidth]{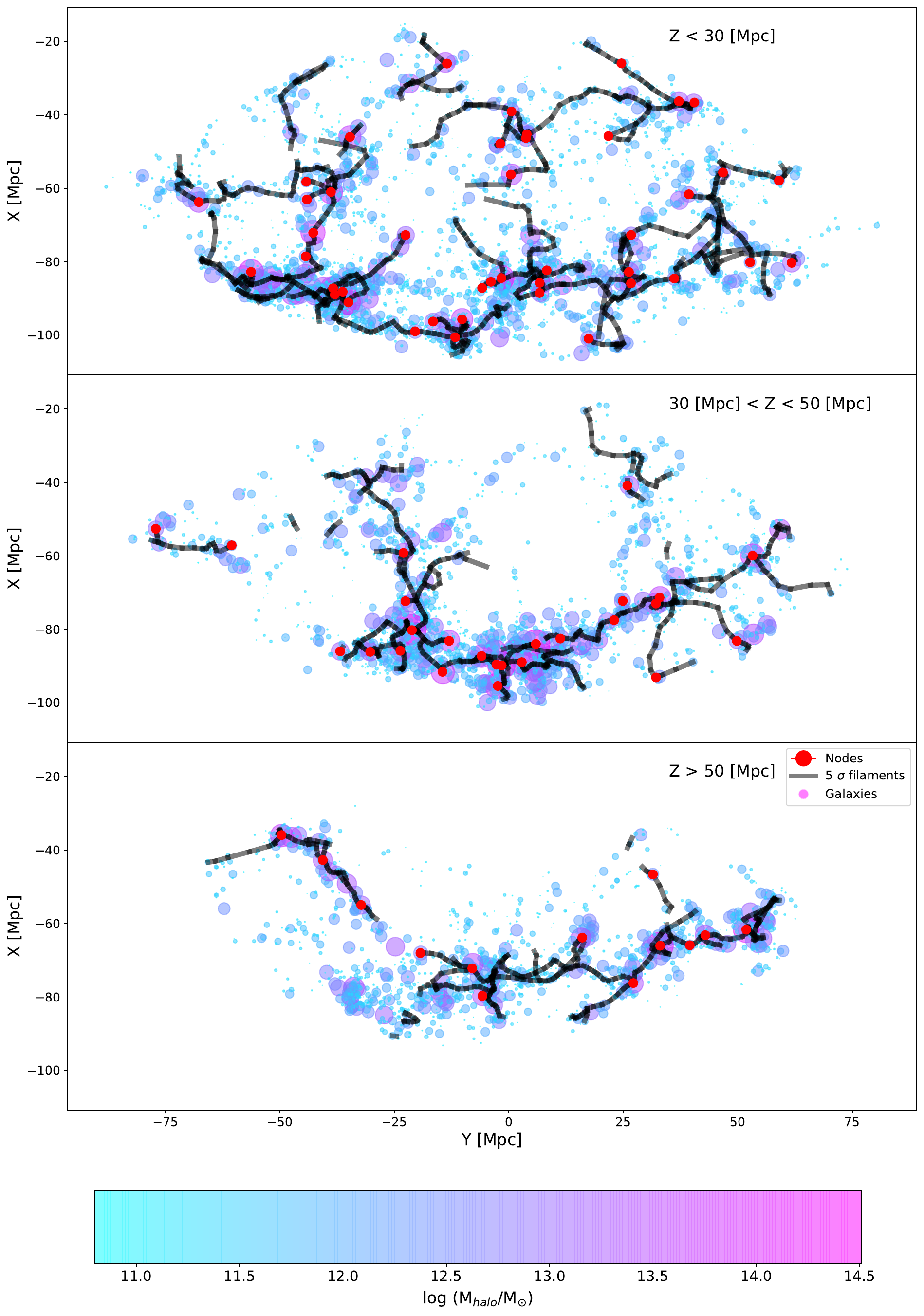}
\caption{Filaments in ECO presented in slices along the Z-axis. The top panel shows the slice where Z < 30 Mpc. The middle panel shows 30 Mpc < Z < 50 Mpc. The lower panel shows Z > 50 Mpc. Galaxies marked as `central' are shown, colour-coded and scaled by their log halo mass. Red circles indicate the position of nodes. }
\label{fig:filaments}
\end{figure*}

We present the filaments identified in the ECO field in Figure~\ref{fig:filaments} where $X$, $Y$, $Z$ refer to the comoving coordinates calculated as per Eq.~\ref{Eq:comoving}. Galaxies identified as `centrals' are also shown in the figure by the circles which are coloured and scaled by their halo masses. The figure clearly shows how the filaments follow the spatial distribution of the high mass haloes.

%\newpage
%\newgeometry{lmargin=2.5cm, tmargin=1.5cm, rmargin=2.5cm, bmargin=2.5cm}
%\begin{figure}
%\end{figure}
%\restoregeometry

Galaxies are divided into bins by their distance to filament (\Dskel) such that each bin has an equal number of galaxies and we calculate the median property in each \Dskel\ bin where applicable. Uncertainty bands are calculated using bootstrapping %\footnote{\texttt{astropy} implementation at https://docs.astropy.org/en/stable/api/astropy.stats.bootstrap.html} 
(1000 iterations with replacement) to determine the 1$\sigma$ confidence interval for each bin when median values are used i.e. when examining trends in stellar mass. The Spearman's rank test %\footnote{\texttt{scipy}  implementation at https://docs.scipy.org/doc/scipy/reference/generated/scipy.stats.spearmanr.html}
is applied to the trends in red fraction and gas-poor fraction to determine their strength and statistical significance.  Trends are considered statistically significant if the p-value is p < 0.003 (more than 3-sigma from null result). The p-values and correlation coefficients for trends in this section can be found in Table \ref{tab:spearmans} in Appendix \ref{appB}.

\subsubsection{Stellar Mass}\label{sec:stellarmass}

\begin{figure}
\includegraphics[width=\columnwidth]{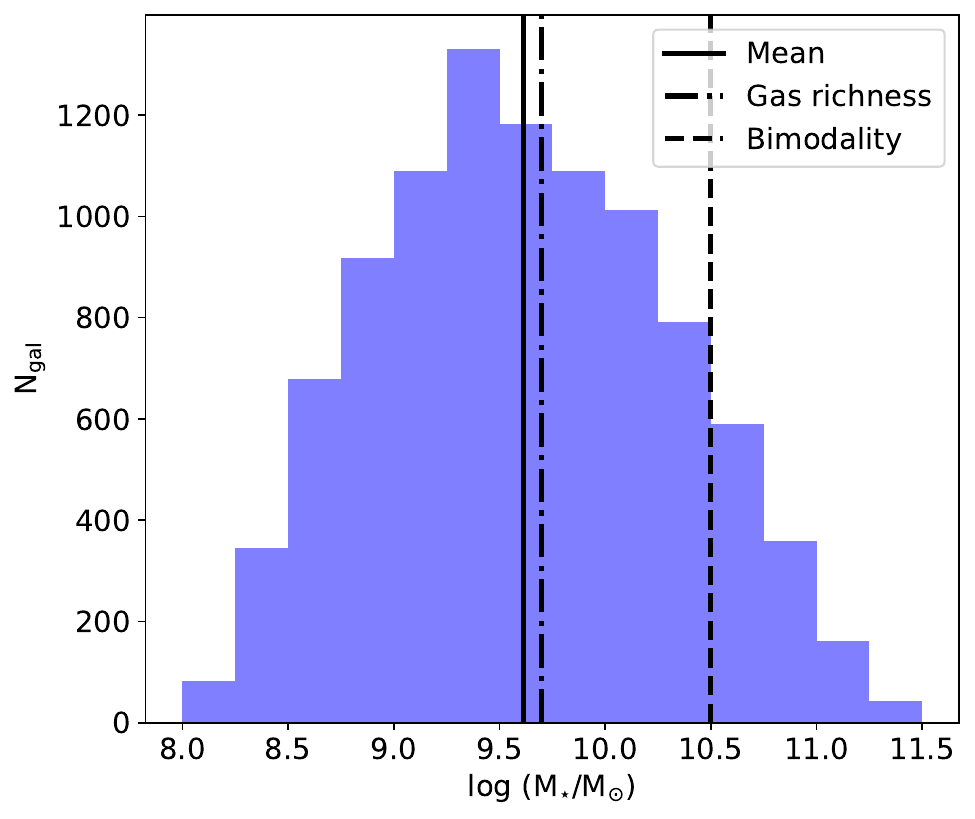}
\caption{Histogram showing the number of galaxies in ECO in stellar mass (log (\mstar/\Msun) bins of 0.5 dex for the baryonic mass selected sample. The mean stellar mass log (\mstar/\Msun )= 9.61 is represented by a solid line. The gas-richness threshold and bimodality scale are indicated by the dotted and dashed lines.}
\label{fig:masshist}
\end{figure}

Stellar mass is one of the primary indicators of galaxy properties \citep{Kauffmann2003a, Alpaslan2015environmentproperties}. ECO includes galaxies across a wide range of stellar masses, as shown in Figure~\ref{fig:masshist}. The histogram shows the distribution of stellar masses for the sample, with the mean stellar mass of log(\mstar/\Msun) =9.61 indicated by the solid line. The gas-richness threshold and bimodality scale, which divide the sample into low, intermediate and high stellar mass sub-samples \citep{kannappan2013gastransitions} (see Section \ref{sec:colour} for details), are shown with dot-dash and dashed lines, respectively. The peak of this histogram is near the gas-richness threshold, indicating that most galaxies in this sample are likely to be gas-rich.

\begin{figure*}
\includegraphics[width=2\columnwidth]{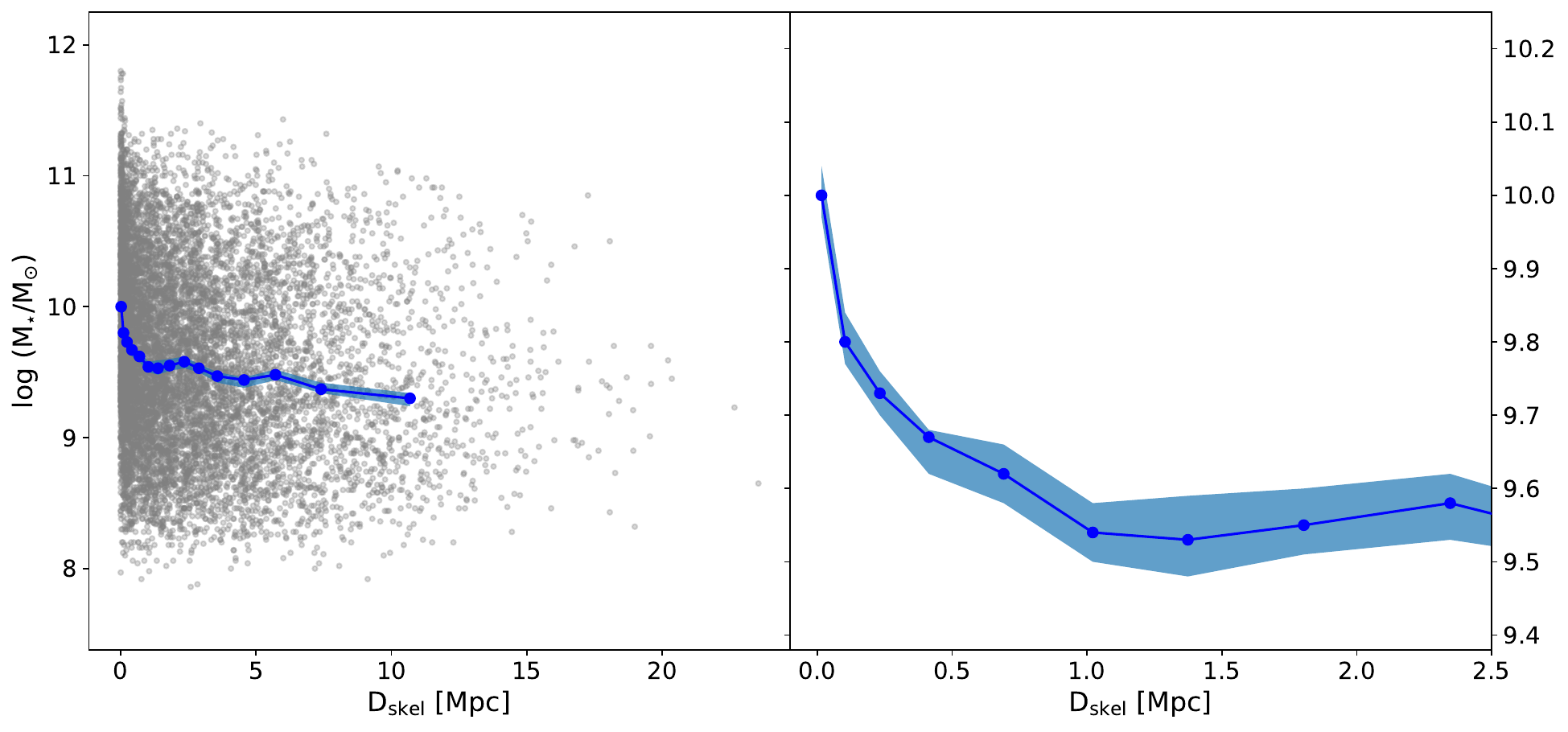}
\caption{Distribution of stellar masses as a function of distance to filament (\Dskel). Individual galaxies are shown in grey. The median stellar mass in each distance bin is shown in blue. The coloured band indicates the 1$\sigma$ error on the median. The right panel shows this distribution for 0 Mpc < \Dskel < 2.5 Mpc and is zoomed-in on the y-axis to highlight the behaviour close to the filaments. The median decreases by 0.5 dex as \Dskel increases within 2 Mpc.}
\label{fig:avemass}
\end{figure*}

In Figure \ref{fig:avemass} we present stellar mass vs distance to filament for the ECO sample (grey points) with the median values overlaid in blue. The coloured band indicates the 1$\sigma$ error on the median in each bin. The right panel shows the trend within 2.5 Mpc of filaments to more clearly display the behaviour at small distances. Note that the y-axis also has a different scale in the right hand panel.

The median stellar mass decreases by 0.5 dex within \Dskel = 2 Mpc and changes by $\sim$ 0.7 dex across the full distance range. We fit a weighted linear function to the median log stellar mass vs log \Dskel using \texttt{scipy}’s curve\textunderscore fit. We find that the slope co-efficient m =-0.095 is statistically different from zero at the 3 $\sigma$ level (uncertainty $\sigma = 0.005$). This indicates that the decrease in the median stellar mass is significant, suggesting that galaxies close to filaments have higher stellar masses than those further away.

\subsubsection{Colour}
\label{sec:colour}

Galaxy colour is an indicator of star formation, quenching, dust content and stellar age. High density regions typically host a higher fraction of red, early-type galaxies than low density regions \citep[e.g.,][]{dressler1980morphologydensity}.

\citet{kannappan2013gastransitions} showed that galaxy populations can be divided into three scales related to their gas refuelling regime: the `accretion-dominated' scale, below the gas richness threshold at log $(M_{\star}/M_{\odot})$~<~9.7 (see \citet{DekelSilk1986}, where quasi-bulgeless galaxies are common and are refuelled by accreting gas; the `process-dominated' regime, where galaxies accerete gas at approximately the same rate at which it is used up by processes such as star formation; and the `quenched' regime, which occurs above the bimodality scale (log $(M_{\star}/M_{\odot})$~>~10.5) \citep{Kauffmann2003a}, and consists of elliptical or S0 galaxies. These scales are used in this study to divide galaxies into stellar mass bins as they segregate galaxies by their type and depend on the gas refuelling regime of a galaxy, ensuring that we can observe trends in galaxy properties due to their environment rather than their stellar mass. We note that this primarily applies to central galaxies within haloes; satellite galaxies may be subject to quenching or refuelling dependent more on their halo properties than their stellar mass. However, this will be further explored in Section \ref{subsec:groupssingles} and Section \ref{sec:environment}. Because the ECO sample contains a wealth of low stellar mass galaxies, we further divide the sample below the gas-richness threshold such that galaxies with log \mstar / \Msun \textless 9.0 are considered `ultra-dwarf'. Although there is no clearly-defined transition for `dwarf' galaxies, this division ensures that any underlying trends with distance to filament due to changes in stellar mass are fully removed. Our stellar mass subsamples are referred to as 'ultra-dwarf' for galaxies with log \mstar/ \Msun \textless 9.0, `low mass' for galaxies with 9.0 \textless log \mstar/\Msun \textless 9.7, `intermediate mass' for galaxies between the bimodality scale and gas-richness threshold, and `high mass' for galaxies above the bimodality scale.   

\begin{figure}
\includegraphics[width=\columnwidth]{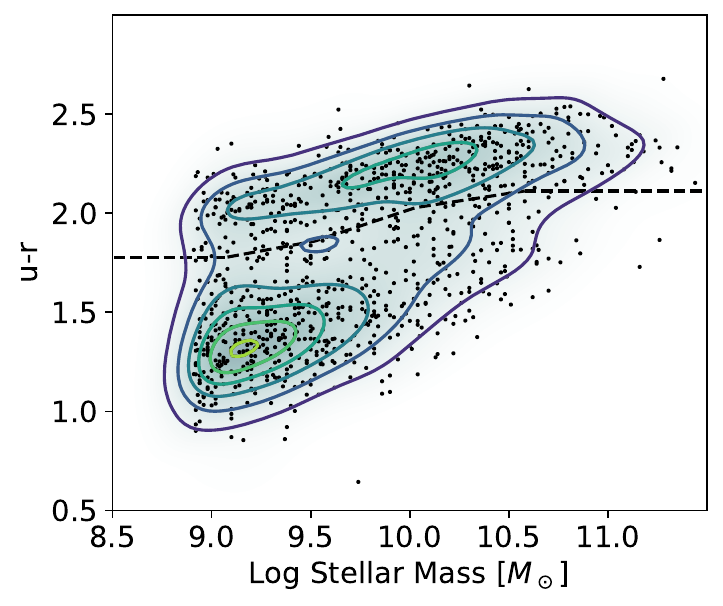}

\caption{The colour-mass diagram for galaxies in the ECO sample. The red vs. blue divider derived from ECO DR3 limited at stellar mass log (\mstar/\Msun) = 8.9.  For visualisation, we plot 1000 random ECO galaxies on top of the Kernel Density Estimate contours derived from the stellar mass-limited ECO sample. }
\label{fig:CMD}
\end{figure}

To define the divider between red and blue galaxies, we performed double gaussian fitting to the distributions of \textit{u-r} colour (foreground extinction and k-corrected) in stellar mass bins using galaxies in the stellar-mass selected sample of the ECO DR3. We select down to a stellar mass limit of log (\mstar/\Msun) = 8.9 for galaxies within group velocities 3500 \kms < $cz_{\mathrm{group}}$ < 7000 \kms to define an approximately complete sample \citep[following][]{eckert2016resolve}, and we set bins from log (\mstar/\Msun) = 8.9 – 9.3, 9.3 – 9.8, 9.8 – 10.3, and above log (\mstar/\Msun) = 10.3. The fitting method is similar to that of \citet{baldry2004CMD}, except we bin in stellar mass instead of magnitude and similar to \citet{Moffett2015ECO}, but with more stellar mass bins. From the fits, we define a point for each bin in which the x-value is the median stellar mass within the bin and the y-value is the \textit{ u-r} colour that marks where the blue-galaxy and the red-galaxy gaussian fits intersect. Finally, we connect each point to create the red-blue divider seen in Figure \ref{fig:CMD}. We note that the divider does not change significantly with $\sim$2x finer binning.

\begin{figure}
\includegraphics[width=\columnwidth]{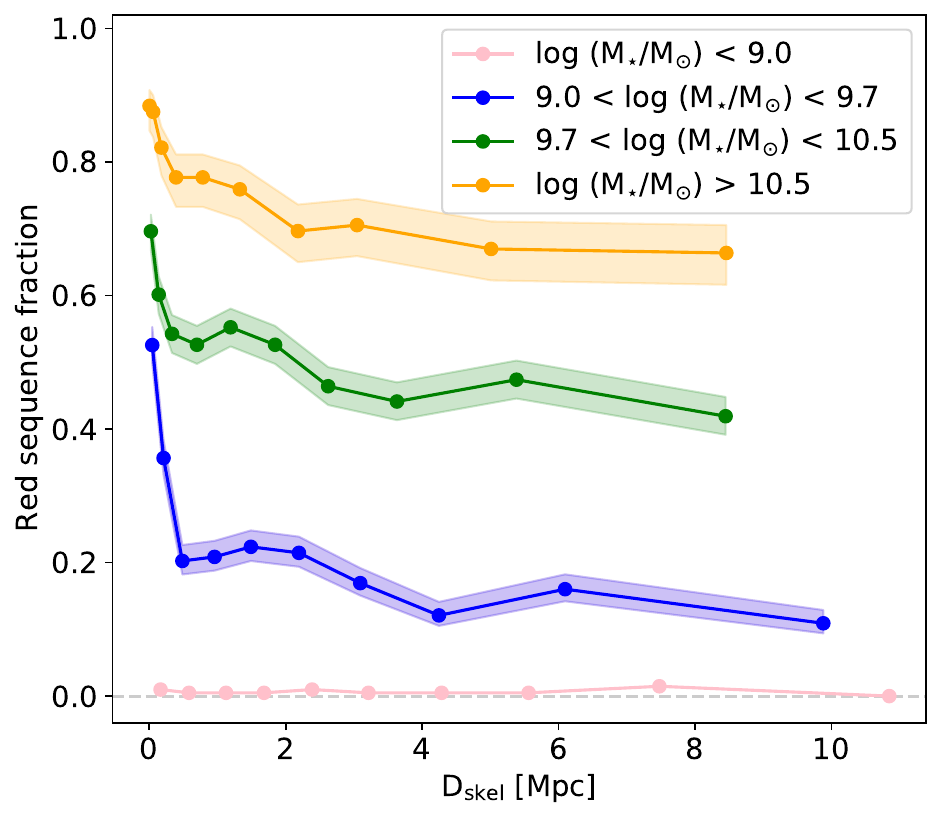}
\caption{The fraction of red sequence galaxies for ultra-dwarf (pink), low mass (blue), intermediate mass (green) and high mass (orange) galaxies vs distance to filament (\Dskel). The red fraction increases close to filaments with statistical significance in each stellar mass bin except for ultra-dwarf galaxies.}
\label{fig:redfrac}
\end{figure}

The fraction of red sequence galaxies (hereafter referred to as the red fraction) in bins of distance to filament for the ultra-dwarf, low, intermediate and high mass sub-samples is shown in Figure~\ref{fig:redfrac}. The coloured bands indicate the 1$\sigma$ uncertainties, which are calculated using Bayesian binomial confidence intervals \citep{cameron2011estimation}. We find a statistically significant increase in the red fraction for galaxies in each stellar mass bin with decreasing distance to filament, with the exception of the ultra-dwarf mass bin, indicating that galaxies are typically redder close to filaments. 

\subsubsection{Gas fraction}
Neutral hydrogen gas is a key component of galaxies and provides the raw fuel for eventual star formation. To examine the possible effects of the filament environment on the gas content of galaxies, we calculate the gas fraction (G/S), defined as 1.4 \mhi/\mstar, where 1.4 \mhi is the atomic hydrogen gas mass including a correction for the helium content. We classify galaxies below a fixed gas fraction (G/S < 0.1) as gas poor rather than using \hi deficiency, which uses the amount of gas relative to that expected as a function of mass or other galaxy properties, since we use photometric gas fractions for many of the ECO galaxies. The photometric fraction makes use of the relationship between optical properties and \hi mass to estimate the fraction of gas expected, thus already taking into account correlations with properties such as mass. We expect this to be a conservative definition of gas-poor for this sample, based on the expected \hi content as a function of mass \citep{Bok2020HIscaling}.

\begin{figure}
\includegraphics[width=\columnwidth]{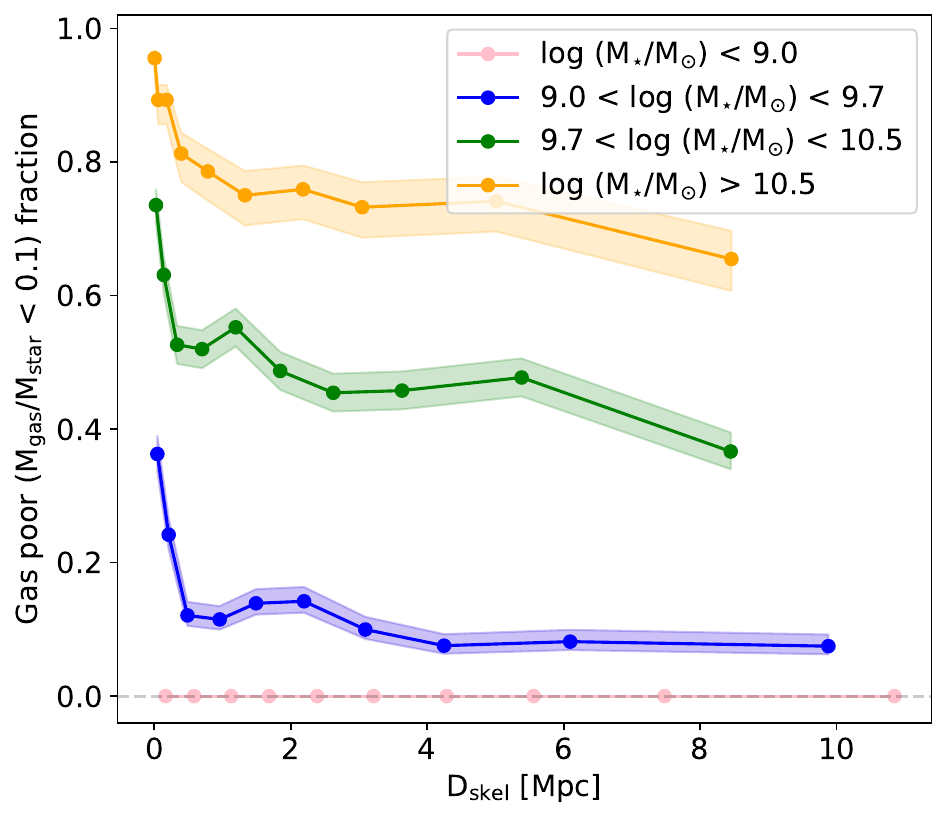}

\caption{The gas-poor fraction vs distance to filament (\Dskel) for ultra-dwarf, low mass, intermediate mass and high mass galaxies. The gas-poor fraction shows a statistically significant increase close to filaments in each stellar mass bin except for ultra-dwarf galaxies.}
\label{fig:gaspoorfrac}
\end{figure}

Figure \ref{fig:gaspoorfrac} shows the fraction of gas-poor galaxies (hereafter referred to as the gas-poor fraction) in bins of distance to filament for each stellar mass sub-sample. As explained in Section \ref{sec:colour}, these stellar mass bins are closely linked to transitions in the gas refuelling regimes \citep{kannappan2013gastransitions}. As such, the low and intermediate mass bins (in the refuelling and processing dominated regimes respectively) are expected to have a higher gas content than the high-mass, quenching-dominated bin. The ultra-dwarf regime has the highest gas content of all the mass bins, and as such, all galaxies within this stellar mass regime are gas-rich, resulting in a gas-poor fraction of zero across the range of distance to filaments. The gas-poor fraction shows a statistically significant increasing trend as one gets closer to filaments in each of the higher stellar mass bins.

Overall, galaxies close to filaments have proportionally less gas than galaxies further away from filaments with the exception of ultra-dwarf galaxies.

\subsection{The effect of groups vs filaments on galaxy properties}
\label{subsec:groupssingles}

Galaxy groups have a significant impact on the evolution of galaxies. Large galaxy groups or clusters  occur at the high density nodes of the cosmic web, which also form the intersections of filaments. These high-density regions may introduce additional gradients and trends in galaxy properties \citep[e.g.][]{Laigle2018cosmos}. Quenching through strangulation or evaporation \citep{fujita2004pre}, may occur in small groups that reside within filaments \citep{Sarron2019preprocessing}. Along with galaxy-galaxy interactions, this may alter the morphology, colour, stellar mass and gas content of galaxies in addition to any effects from the filaments themselves. To isolate the effect of filaments, galaxy groups must be carefully considered. 

Many galaxy groups occur close to the filaments, at distances where trends in stellar mass, colour and gas fraction were observed in Section \ref{sec:galprop}. In the following sections we will consider the effect of filaments on galaxies' properties taking into account whether the galaxies are in groups or isolated within their haloes.\citet{Hutchens2023G3} produced a novel `Gas in Galaxy Groups' (G3) group finder which they applied to the RESOLVE and ECO data. This group finder identifies groups in a four-step process in which giant galaxies are first identified and groups are formed accordingly, with an iterative process assigning galaxies to these giant groups before identifying 'dwarf only' groups. This technique is powerful when dealing with incomplete samples, as dwarf galaxies are more difficult to detect in surveys than larger galaxies. However, because the analysis presented in this paper uses the highly complete ECO and RESOLVE surveys, we use the friends-of-friends group catalogue. We also note that the friends-of-friends groups are less complete in the high-mass regime. However, this is implicitly taken into consideration in our analysis when examining the effect of nodes (Section \ref{sec:nodes}).

\begin{figure}
\includegraphics[width=\columnwidth]{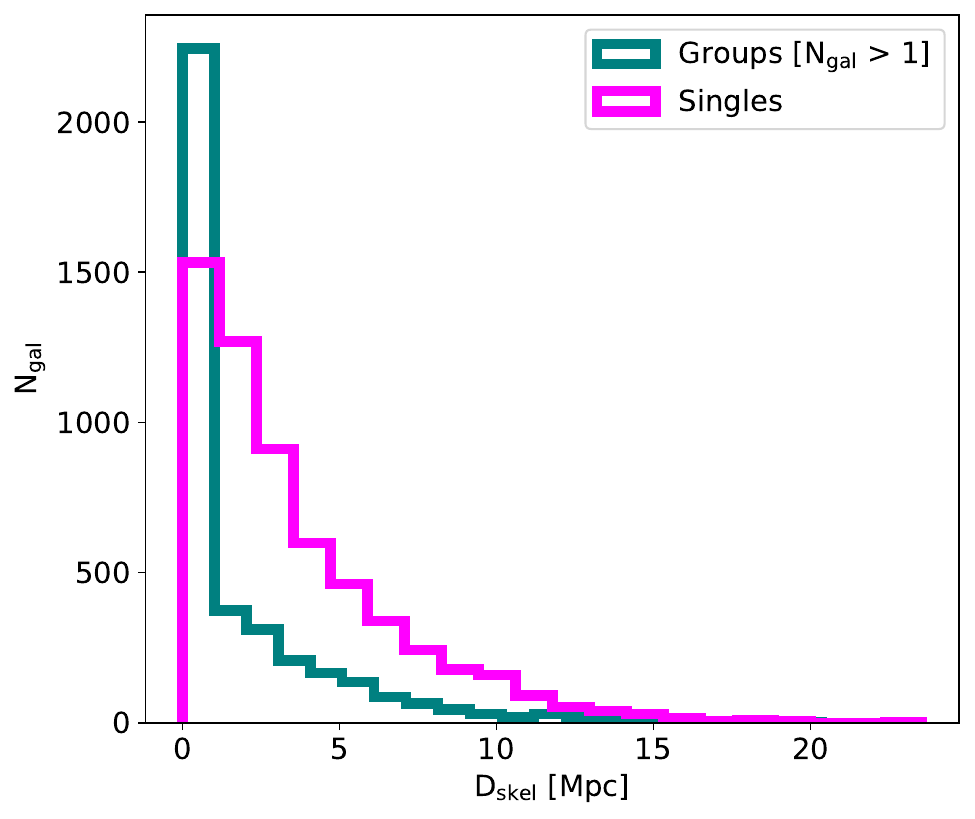}
\caption{The distribution of single galaxies, and galaxies in groups, with distance from filament. Group galaxies are clustered at low \Dskel, while single galaxies are spread out over larger distances.}
\label{fig:grouppos}
\end{figure}

Using the group catalogue information from ECO, we separate our sample into single galaxies and those in groups of more than one galaxy (N>1). Note that while we refer to `single' galaxies in this study, it is likely that these galaxies are not truly \textit{isolated}. Rather, it is possible that these galaxies may have companions that are below the detection floor. For the purpose of this work in differentiating the effect of filaments and large galaxy groups on galaxy properties, it is sufficient to treat these galaxies as `single'. Figure~\ref{fig:grouppos} shows the distribution of group galaxies and single galaxies with distance to filament. Galaxy groups are clustered at low \Dskel, close to filaments, while single galaxies are spread over the full distance range. To isolate the effect of filaments, we use the ECO group catalogues to separate out the the effects of high density regions rather than removing galaxies close to `node' regions identified by \texttt{DisPerSE} as in other works. By isolating single galaxies for this analysis, we remove the effect of clustering in the `node' regions as well as any additional effects due to the group environment, to look at the effects due to filaments alone. A detailed discussion of this is presented in Section \ref{sec:nodes} and Appendix \ref{appNodes}.

\subsubsection{Stellar Mass}

\begin{figure*}
\includegraphics[width=2\columnwidth]{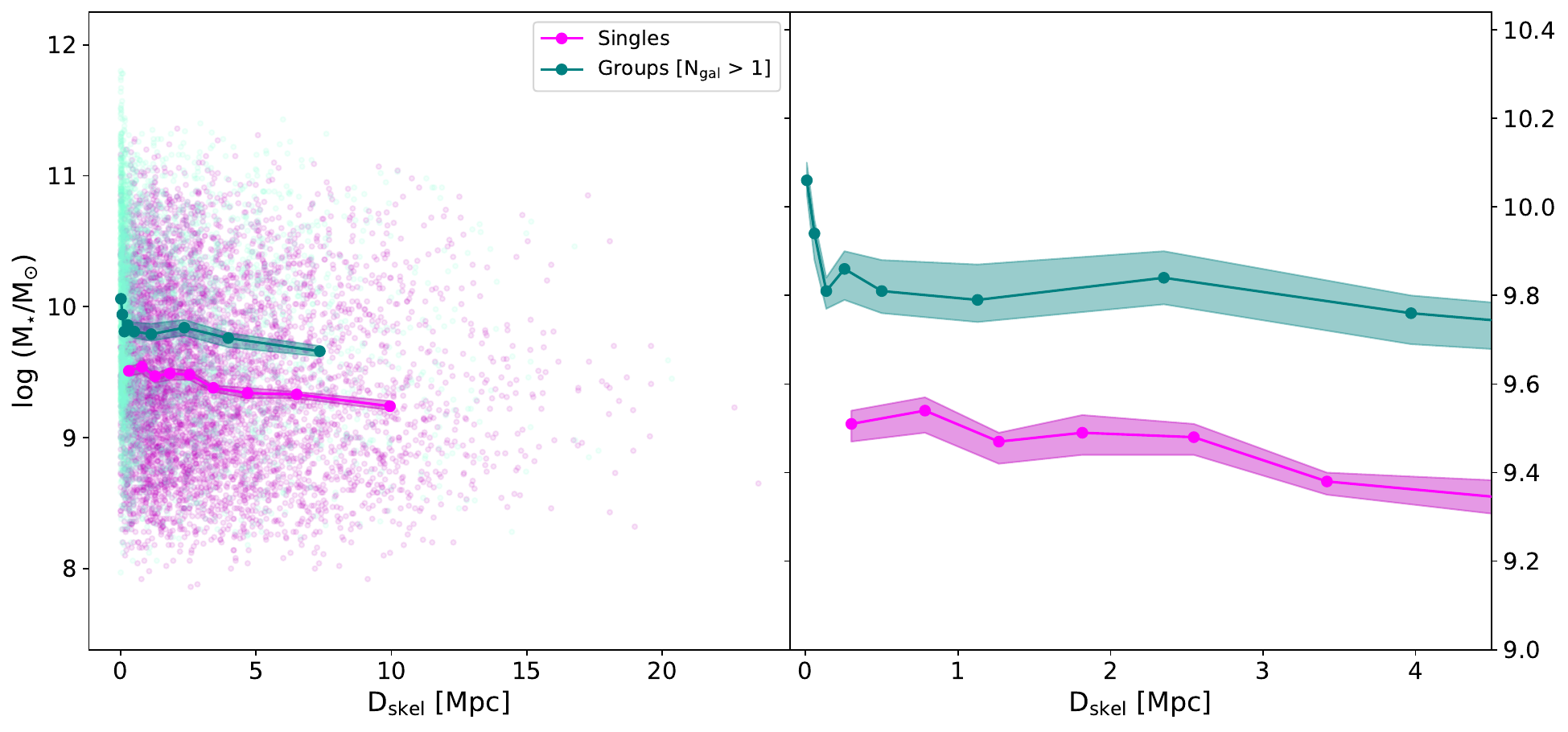}
\caption{The median stellar mass vs distance to filament (\Dskel) for galaxies in groups (teal) and single galaxies (magenta).  Galaxies in groups have systematically higher stellar masses than single galaxies. The right panel shows this distribution for 0 Mpc < \Dskel < 4 Mpc and the 9.0 < (\mstar/\Msun) < 10.4 range on the y-axis to highlight the behaviour close to the filaments.}
\label{fig:mstargroupssingles}
\end{figure*}

In Section~\ref{sec:stellarmass}, we showed that galaxies located closer to filaments tend to have higher median stellar masses. This mass segregation is prominently observed in the literature \citep[e.g.][]{kraljic2018, Laigle2018cosmos, Luber2019CHILES}. To examine the role of galaxy groups in driving this trend, Figure \ref{fig:mstargroupssingles} shows the median stellar mass for group galaxies and single galaxies. Over the full range of \Dskel, the group galaxies have systematically higher stellar masses than the single galaxies. A weighted linear fit was performed to the log stellar mass and log distance to filament for both single and group galaxies, with a statistically significant slope m$_\mathrm{group}$ = -0.050 ($\sigma_\mathrm{group}$ = 0.008) and m$_\mathrm{single}$ = -0.080 ($\sigma_\mathrm{single}$ = 0.016) for both single and group galaxies. This indicates that both group and single galaxies have higher stellar masses close to filaments. 

\subsubsection{Colour}
\label{sss:colour}

\begin{figure}
\includegraphics[width=\columnwidth]{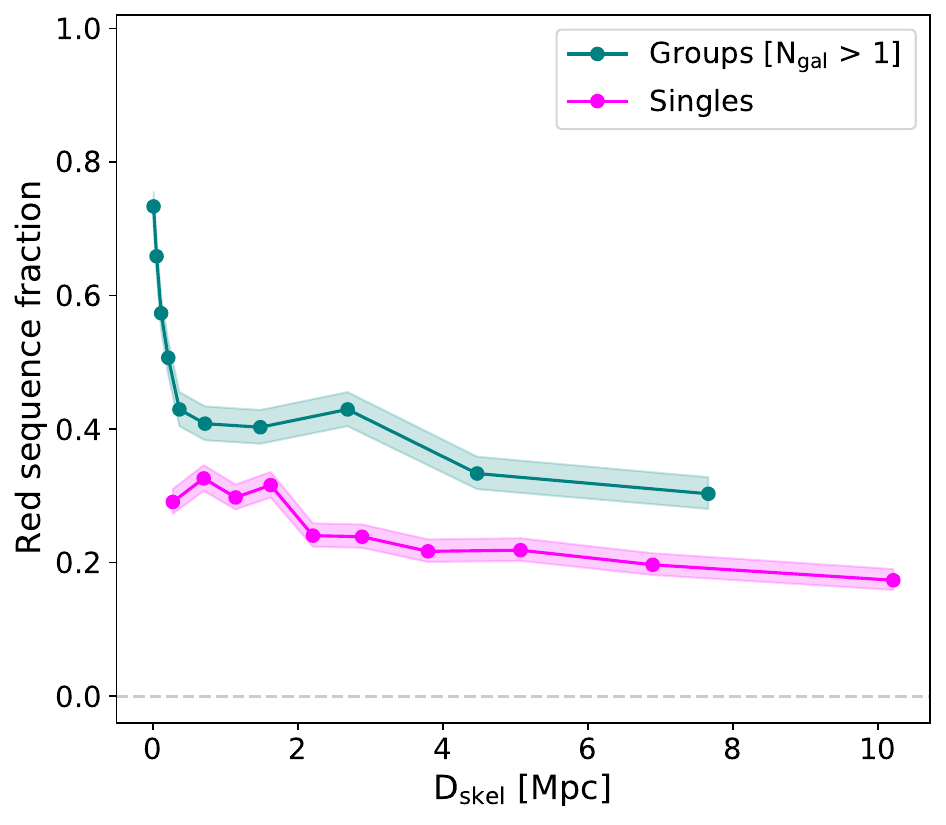}
\caption{The red fraction vs distance to filament for galaxies in groups (teal) and single galaxies (magenta). Both sub-samples show an increase in red fraction as distance to filament decreases. }
\label{fig:redfracgroupssingles}
\end{figure}

As shown in section \ref{sec:colour}, the red fraction and median colour suggest that galaxies are typically redder close to filaments. Higher red fractions and redder colours close to filaments have been found in the literature previously \citep[e.g.][]{Chen2017filaments, kraljic2018} However, group pre-processing is known to quench star formation and induce reddening \citep{davies2019groupquenching, peng2010mass}. To investigate if the observed reddening can be directly attributed to the filament environment, Figure \ref{fig:redfracgroupssingles} shows the red sequence fraction for single galaxies and galaxies in groups vs distance to filament separately. For both single galaxies and group galaxies, the red sequence fraction increases with proximity to the filament. 

The group galaxy trend is very steeply increasing for \Dskel<~0.1 Mpc compared to the trend for single galaxies. 
However, Figure~\ref{fig:mstargroupssingles} showed that group galaxies have higher stellar masses over the full range of \Dskel\ and it is therefore important to ensure that the colour trend is not a stellar mass effect.

\begin{figure*}
\includegraphics[width=2\columnwidth]{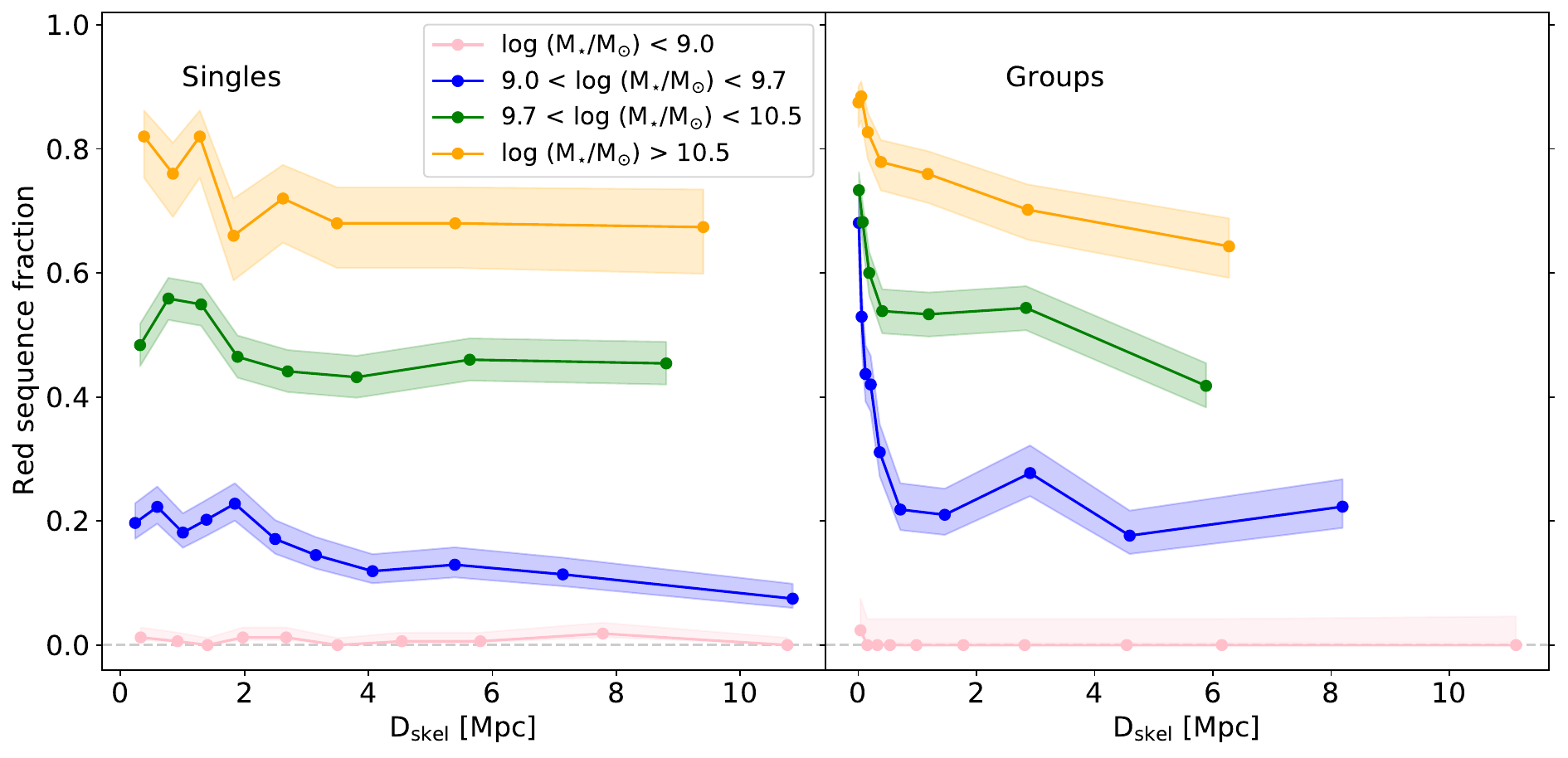}
\caption{The red fraction vs distance to filament is shown for single galaxies (left panel) and group galaxies (right panel) for ultra-dwarf, low, intermediate and high stellar mass sub-samples. Group galaxies show strong, statistically significant increases in red fraction with decreasing \Dskel for all mass bins. Statistically significant increasing trends are also seen for the low mass single galaxies. }
\label{fig:redfraccomp}
\end{figure*}

Figure \ref{fig:redfraccomp} shows the red fraction for group and single galaxies in bins of stellar mass as a function of \Dskel. The increase in red fraction with decreasing distance to filament for single galaxies is statistically significant for the low stellar mass bin. 
In group galaxies, there is a significant increase in red fraction for all stellar mass bins. A small increase in the red fraction close to filaments is present in the ultra-dwarf mass regime. However, this stellar mass bin has very few red galaxies due to the inherent blue colour of ultra-dwarf galaxies. This implies that filaments have a small effect on the colour of low mass galaxies, and reddening is primarily driven by galaxy groups close to filaments.

\subsubsection{Gas fraction}

\begin{figure}
\includegraphics[width=\columnwidth]{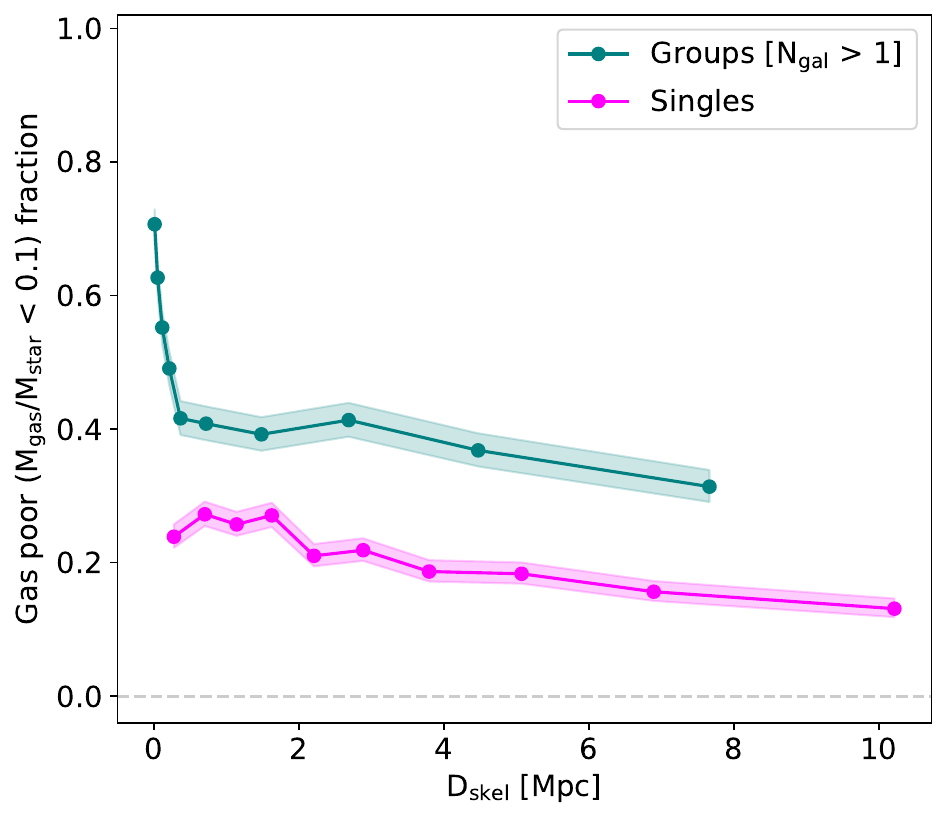}
\caption{Gas-poor fraction vs distance to filament for group galaxies (teal) and single galaxies (magenta). Both sub-samples show a significant decrease in the gas-poor fraction as \Dskel increases. }
\label{fig:gaspoorfracgroupssingles}
\end{figure}

To investigate the relative effects of galaxy groups vs the filament environment on the gas content of galaxies, the fraction of gas-poor single galaxies is compared to the fraction of gas-poor group galaxies as a function of \Dskel\ in Figure \ref{fig:gaspoorfracgroupssingles}. 
Single galaxies and galaxies in groups both show a statistically significant increase in the fraction of gas-poor galaxies as distance to filament decreases. However, galaxies in groups are more gas-poor overall.

\begin{figure*}
\includegraphics[width=2\columnwidth]{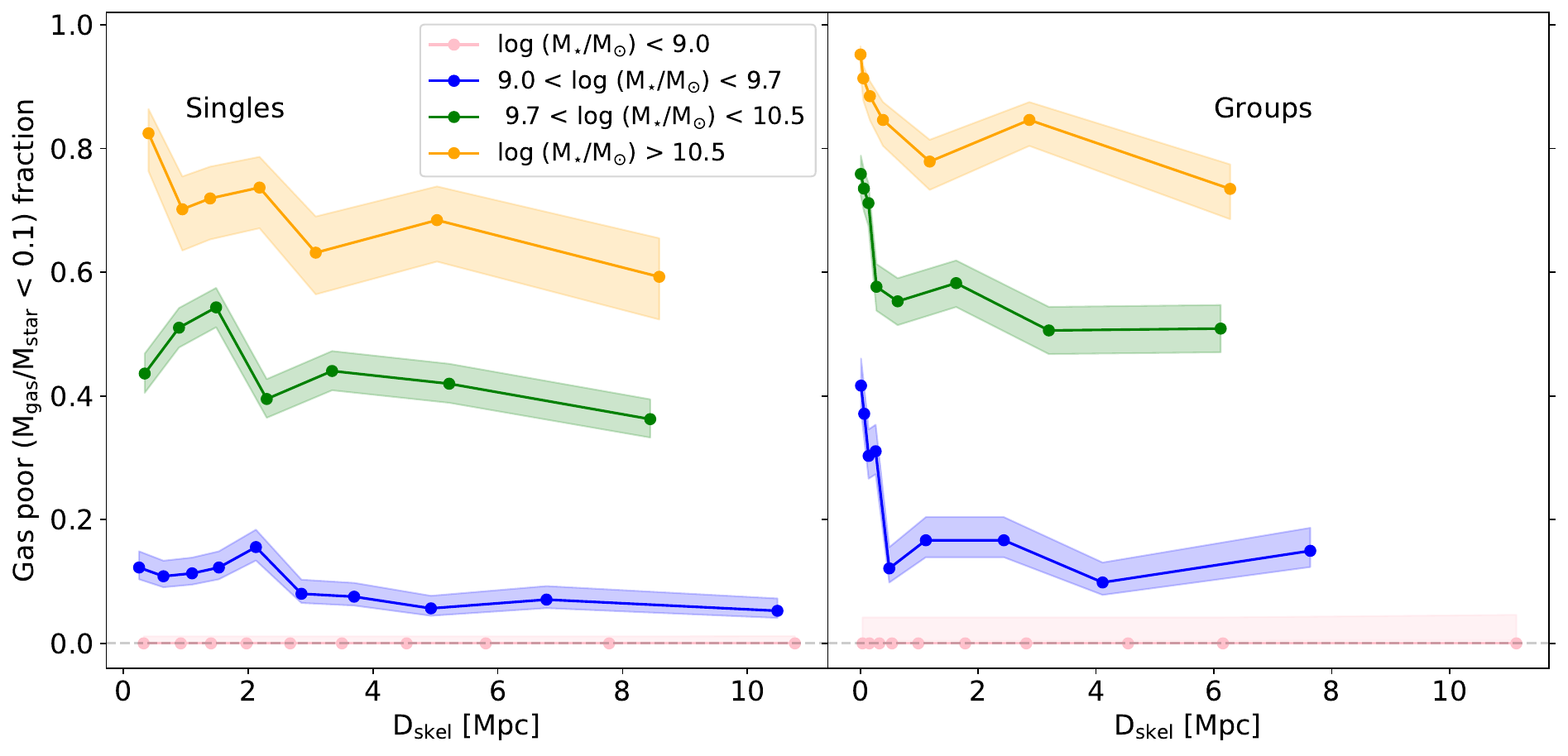}
\caption{Gas-poor fraction vs distance to filament for single galaxies (left panel) and group galaxies (right panel) for ultra-dwarf, low, intermediate and high stellar mass sub-samples. Low and intermediate mass group galaxies show statistically significant increases in gas-poor fraction with decreasing \Dskel. A statistically significant increase in the gas-poor fraction of low-mass, single galaxies is also found.}. 
\label{fig:gaspoorcomp}
\end{figure*}

When further broken down by stellar mass, similarly to Section \ref{sss:colour} and shown in Figure \ref{fig:gaspoorcomp}, we see a statistically significant increase in the gas-poor fraction as distance to filament decreases for low and intermediate mass group galaxies. We find a small, statistically significant increase in the gas-poor fraction close to filaments for high-mass, single galaxies, and no significant trends in the remaining mass bins for single galaxies. This implies that changes in gas fraction are more closely related to group membership than proximity to filaments.  %We do not find any statistically significant trends in gas-poor fraction for single galaxies in these stellar mass bins. %\textcolor{red}{Because galaxies with higher stellar masses are typically found in groups close to filaments, it is unsurprising that stellar mass dominates over distance to filament in two dimensional fits to the gas-poor fraction for group galaxies in all mass bins (see Table \ref{tab:c0c1values}). }

\section{Discussion} \label{sect:discussion}

The current framework for understanding how filaments affect galaxy evolution suggests that the filament `backbone' is made up of higher-mass galaxies and groups. This has been supported by observations of mass segregation within filaments in simulations and data - where galaxies at the core of filaments have higher stellar masses than galaxies outside filaments \citep[e.g.][]{Chen2017filaments, kraljic2018}. Furthermore, Tidal Torque Theory \citep{hoyle1949TTT,Peebles1969TTT, Doroshkevich1970TTT, White1984TTT} provides a mechanism for low-mass galaxies to accrete gas at the `vorticity-rich' outskirts of filaments \citep{Laigle2015TTT}. Additionally, as galaxies enter filaments, they may become detached from their primordial gas supply through major mergers, accretion of satellites or as they cross the filament.  Filaments are typically classified as intermediate-density regions. As such, various authors \citep[for e.g.][]{Guo2015Satellites,Kuutma2017filaments,AragonCalvo2019CWD} suggest that galaxy-galaxy mergers and interactions may occur frequently in filaments, driving morphological transformations such as those observed by \citet{Kuutma2017filaments}. Galaxies in groups travelling along filaments to higher-density regions such as clusters may be pre-processed by the groups, which may result in morphological and gas fraction changes to the galaxies \citep{fujita2004pre,Sarron2019preprocessing}.

In the previous section, we showed that galaxies close to filaments have higher stellar masses, are redder and are more gas-poor than galaxies further away and that these trends generally still hold when the sample is divided into stellar mass bins. However, we found that group environments within filaments may be responsible for these trends.

In this section, we examine the possible mechanisms that may drive these trends in groups and filaments and compare the results to previous work.

\subsection{Trends in Stellar Mass}

It is well established that stellar mass is a crucial predictor of a galaxy's properties, even when environment is considered \citep{kauffmann2004environmental,Kauffmann2003a, peng2010mass, Alpaslan2014b}. In the previous section, Figure \ref{fig:avemass} showed that the median stellar mass of galaxies is higher closer to filaments than further away.

\begin{figure*}
\includegraphics[width=2\columnwidth]{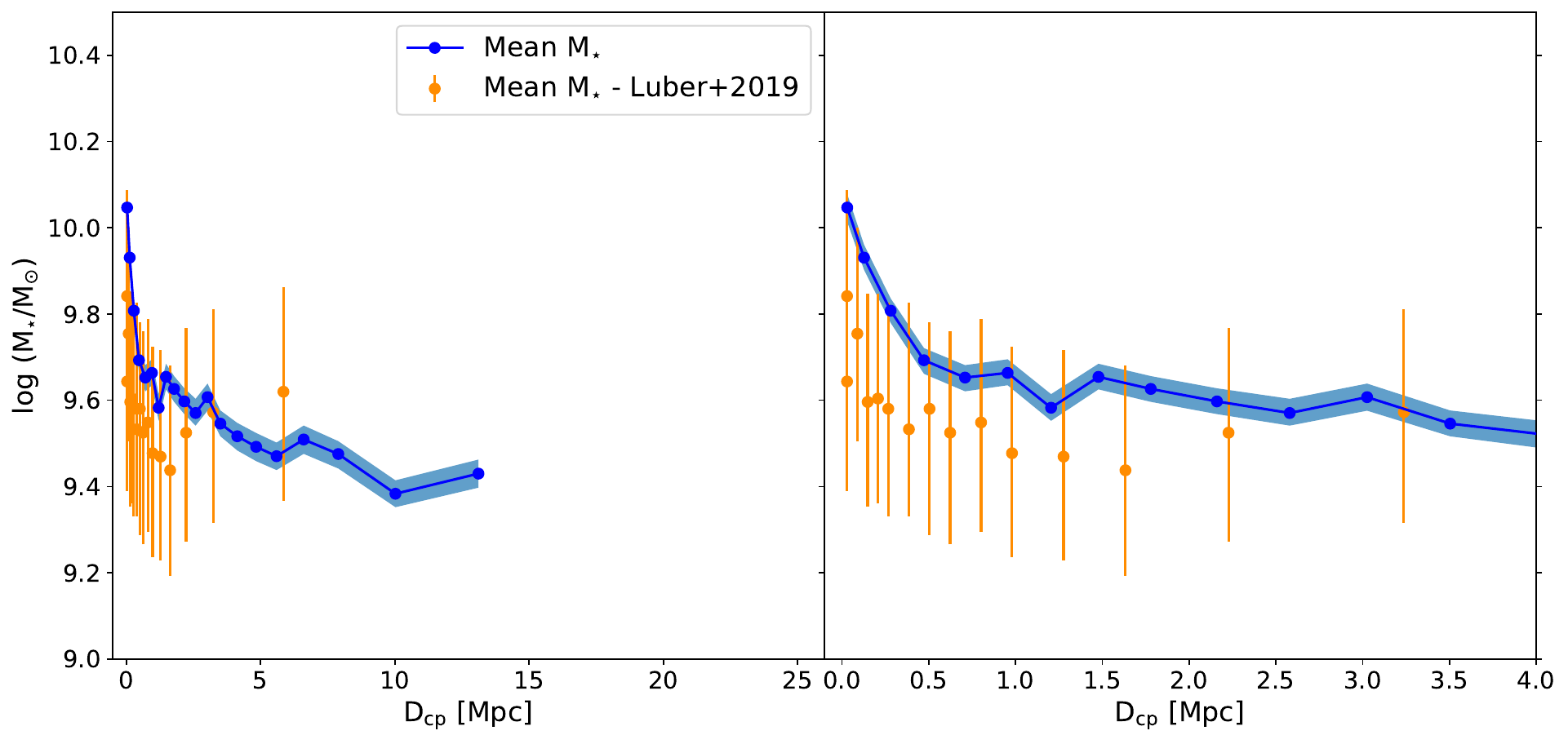}
\caption{The mean stellar mass vs distance to nearest critical point (D$_\mathrm{cp}$ for ECO (blue) is shown in comparison to the same parameters and errorbars from \citet{Luber2019CHILES} (orange). The uncertainty band for the ECO data is calculated as previously (see Section \ref{sect:results}) The right-hand panel shows this for D$_\mathrm{cp}$ < 4 Mpc to emphasise the behaviour close to the critical points. }
\label{fig:avemasscompluber}
\end{figure*}

While \Dskel, defined as the transverse distance to the filament, is most commonly used as the metric for measuring the distance to filaments in studies that use \texttt{DisPerSE} for filament identification, \cite{Luber2019CHILES} and \cite{blue2020chiles} use the distance to nearest critical point, which we consider as the end points of filament segments, D$_{\mathrm{cp}}$, in their analyses of galaxy properties. As described in Section~\ref{sec:FilDist}, D$_{\mathrm{cp}}$ was also calculated for each galaxy. For comparison to \citet{Luber2019CHILES}, we present the mean stellar mass, calculated in bins of D$_{\mathrm{cp}}$, in Figure \ref{fig:avemasscompluber}. This figure shows that in the ECO data the mean stellar mass increases with decreasing distance to critical point. Although the ECO data are offset at higher stellar masses overall than the data from \citet{Luber2019CHILES}, possibly due to differences in the sample selection and the use of mean values to represent the binning, both data sets follow the same trend. Within the uncertainties given by the dispersions on stellar mass values (refer to Figure~\ref{fig:avemass}) per D$_{\mathrm{cp}}$ bin, our results are consistent with those from \citet{Luber2019CHILES}.

In this work, we primarily separate galaxies into single and group galaxies (see Section \ref{sect:results}) to account for the variation of environment and density within filaments.
Single galaxies close to filaments are found to have higher median stellar masses than galaxies further away (see Figure~ \ref{fig:mstargroupssingles}). This effect is small (< 0.5 dex), but statistically significant with the slope = 0.08 and the uncertainty $\sigma$ = 0.014) and occurs within \Dskel < 2.5 Mpc. This increase is in agreement with work by many authors \citep[e.g.][]{Alpaslan2015environmentproperties, kraljic2018, Luber2019CHILES}  and suggests that mass segregation occurs within filaments, regardless of whether a galaxy is in a group or isolated in its halo.  Note that single galaxies as we have defined them are not necessarily isolated, but are single relative to our selection floor and may have lower mass companions. For the lowest mass bins, single galaxies could be part of dwarf only groups, whereas the "single" galaxies in higher mass bins have stellar masses well above any undetected satellites, and could be considered isolated from any similar mass neighbours.

\subsection{Trends in Colour}

The colour-magnitude diagram reveals that a sample of galaxies will typically form a bimodal distribution - a red sequence consisting of quenched, mostly early-type galaxies, and a `blue cloud' of galaxies that are actively forming stars \citep{baldry2004bimodalcolourmag}. 

Colour is also closely tied to environment and galaxy density. In high-density regions such as galaxy clusters, the red sequence dominates \citep{hogg2003overdensities}. Filaments may be a site for pre-processing as galaxy groups travel towards clusters, which may redden galaxies in groups by quenching star-formation through mechanisms such as strangulation and evaporation, reducing the quantity of gas available to form stars \citep{fujita2004pre}. In addition, recent studies found evidence for an enhanced red fraction \citep{kraljic2018, Chen2017filaments}, redder colour \citep{Laigle2018cosmos, Kuutma2017filaments, Luber2019CHILES} and an increased fraction of passive (i.e. non star-forming) galaxies close to filaments \citep{kraljic2018,malavasi2017vimos, Sarron2019preprocessing, Laigle2018cosmos} suggesting that quenching mechanisms may be at play within filaments. 

Our results appear to agree with these findings and hold for all bins in stellar mass, except for ultra-dwarf galaxies, as shown in Figure ~\ref{fig:redfrac}. However, when we divide our sample into group galaxies and single galaxies, we find that these trends are driven predominantly by galaxy groups (see Figures~\ref{fig:redfracgroupssingles} and\ref{fig:redfraccomp} ).  
We find that the trend in red fraction is primarily driven by the group environment rather than the filament environment. Some reddening close to filaments is observed for low-mass, single galaxies. \citet{Malavasi2022cosmicweb} found that star formation rates strongly depend on local environment, which is echoed in our results showing that colour is affected by the group environment significantly more than the filament environment. Additionally, they found that trends in galaxy properties with respect to filaments are more evident in low-mass galaxies, in agreement with our findings.

\subsection{Trends in Gas Fraction}

While trends in colour and stellar mass with respect to filaments have been well-established in the literature, understanding the role of filaments on the gas content of galaxies requires more investigation. Theoretical work has established Tidal Torque Theory to describe how angular momentum flows occur and transfer to galaxies within the cosmic web  \citep{hoyle1949TTT, Peebles1969TTT, Doroshkevich1970TTT, White1984TTT,Porciani2002SpinTTT,codis2015SpinTTT}. Vorticity-rich regions at the outer edges of filaments may allow low-mass galaxies to accrete cold gas efficiently by having their spin (angular momentum) parameters aligned with close-by filaments \citep{Laigle2015TTT, Laigle2018cosmos}. Observations and simulations have attempted to link this spin alignment or mis-alignment with the stellar mass and \hi\ mass of galaxies. \citet{Welker2019spin} found a stellar-mass dependence, with galaxies log (\mstar/\Msun) < 10.4 spin-aligned to their host filaments and a transition mass at  10.4 < log (\mstar/\Msun) < 10.9 where galaxies no longer had their spin aligned. This was also tentatively observed by \cite{blue2020chiles} using a small sample from the CHILES survey. \citet{kraljic2020spin} used data from the SIMBA simulations \citep{dave2019simba} and found that galaxies with high \hi  masses (log (\mhi /\Msun) > 9.5) had their spin aligned to nearby filaments and galaxies with low \hi\ mass (log (\mhi /\Msun) < 9.5) had perpendicular spin to their filaments, further showing the link between stellar mass, gas accretion and the cosmic web. \citet{Song2021filaments} carefully considered the positions and angular momentum of galaxy haloes with respect to filaments. In agreement with \citet{Laigle2015TTT}, they described that haloes in the vorticity rich outskirts of filaments accrete matter due to their angular momentum alignment and that galaxies in this region may be susceptible to additional quenching. However, transferring this matter to galaxies residing in these haloes is an inefficient process which may not translate into changes in the galaxy properties.

Observationally, \citet{kleiner2016DisPerSE} found that high mass (log~(\mstar /\Msun)>11) galaxies close to filaments, with \Dskel~<~0.7~Mpc, had higher gas fractions than their control sample with \Dskel > 5 Mpc. They interpreted this as possible \hi\ cold accretion by massive galaxies from filaments. This mass range is higher than the transition mass found in studies of Tidal Torque Theory \citep{Welker2019spin,kraljic2020spin}, which indicates that more work is needed in this area of research. The ECO sample contains only 24 single galaxies with log(\mstar /\Msun) > 11. Due to these limited statistics, it is not possible to compare this work directly to the results from \citet{kleiner2016DisPerSE}. 

\begin{figure}
\includegraphics[width=\columnwidth]{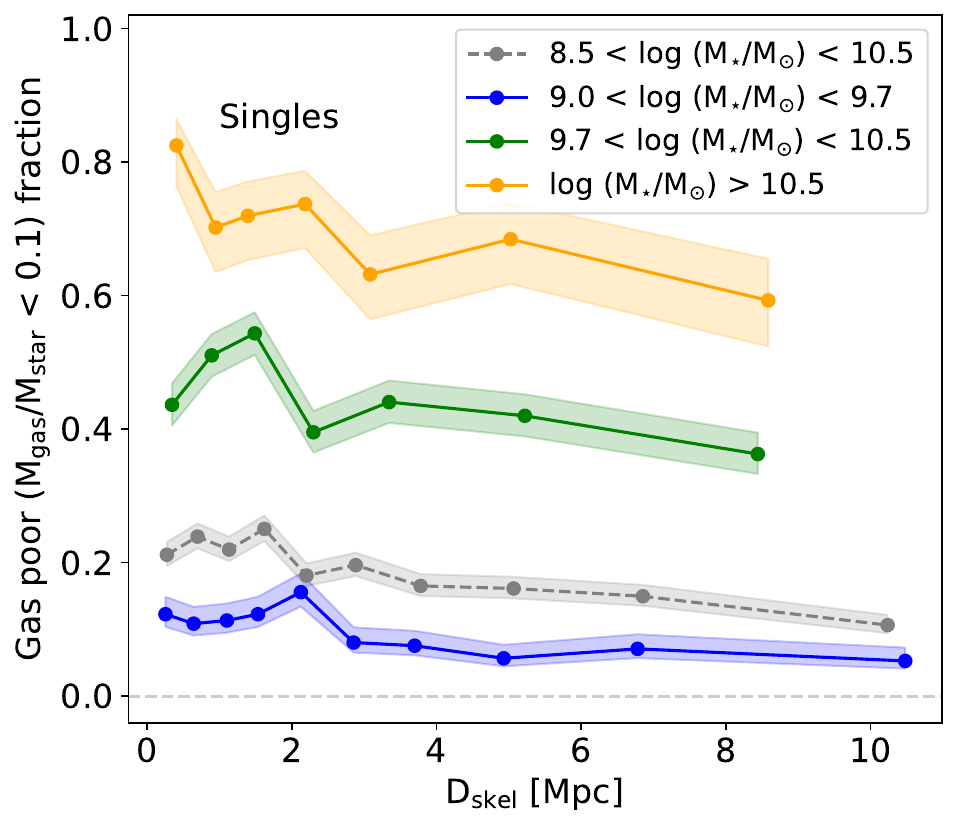}
\caption{The gas-poor fraction vs distance to filament for single galaxies in the previously defined high, intermediate and low mass bins are plotted in orange, green and blue solid lines. In comparison to \citet{odekon2018effect}, the gas-poor fraction for galaxies with 8.5 < log (\mstar/\Msun) < 10.5 is shown with a dashed grey line. The gas-poor fraction for galaxies in this mass range shows a statistically significant increase as distance to filament decreases. }
\label{fig:gaspoorodekoncomp}
\end{figure}

On the other hand, \cite{odekon2018effect} found that at fixed stellar masses and colour, filament galaxies, with 8.5 < log (\mstar/\Msun) < 10.5, are more \hi\ deficient than non-filament galaxies. They suggested a scenario where galaxies enter filaments and are cut off from their gas supply, resulting in the observed \hi\ deficiencies, and later redden as star formation is quenched. This is in agreement with expectations from the Cosmic Web Detachment model \citep{AragonCalvo2019CWD}, which describes how galaxies are quenched after being detached from their primodial gas supply when entering the cosmic web. To compare to \citet{odekon2018effect}, we plot the gas-poor fraction for single galaxies with 8.5 < log (\mstar/ \Msun) < 10.5 with a grey dashed line in Figure \ref{fig:gaspoorodekoncomp}. Single galaxies are selected as \citet{odekon2018effect} also removed the effect of groups in their results. We observe a statistically significant increasing trend in the gas-poor fraction close to filaments within this mass range. This indicates that galaxies in this mass range have less gas close to filaments, in agreement with \citet{odekon2018effect}. Galaxies within this mass range fall within the `process-dominated' and `accretion-dominated' regimes described by \citet{kannappan2013gastransitions}, indicating that these galaxies should be typically gas-rich. Thus, an increase in the gas-poor fraction close to filaments within this mass regime shows that filaments may result in a small reduction in the gas content of galaxies.

Although we and \citet{odekon2018effect} detect a decrease in the gas content of galaxies and \citet{kleiner2016DisPerSE} detect an increase in the gas content of galaxies due to the cosmic web, these changes are observed in different stellar mass regimes. One interpretation of this could be that galaxies with very high stellar masses may have large enough gravitational potentials to funnel gas from the cosmic web \citep{kleiner2016DisPerSE}, while low mass galaxies are more susceptible to `cosmic web stripping' which removes gas from galaxies through ram-pressure inside filaments \citep{Benitez-Llambay2013cosmicwebstripping}, and to Cosmic Web Detachment \citep{AragonCalvo2019CWD} where they are cut off from their primordial gas supply once they enter filaments. Evidence of cosmic web stripping was also observed by \citet{Winkel2021quenching} using data on the sSFR and metallicity of galaxies in the SDSS.

\subsection{Environments and gas content}
\label{sec:environment}

\begin{figure*}
\includegraphics[width=2\columnwidth]{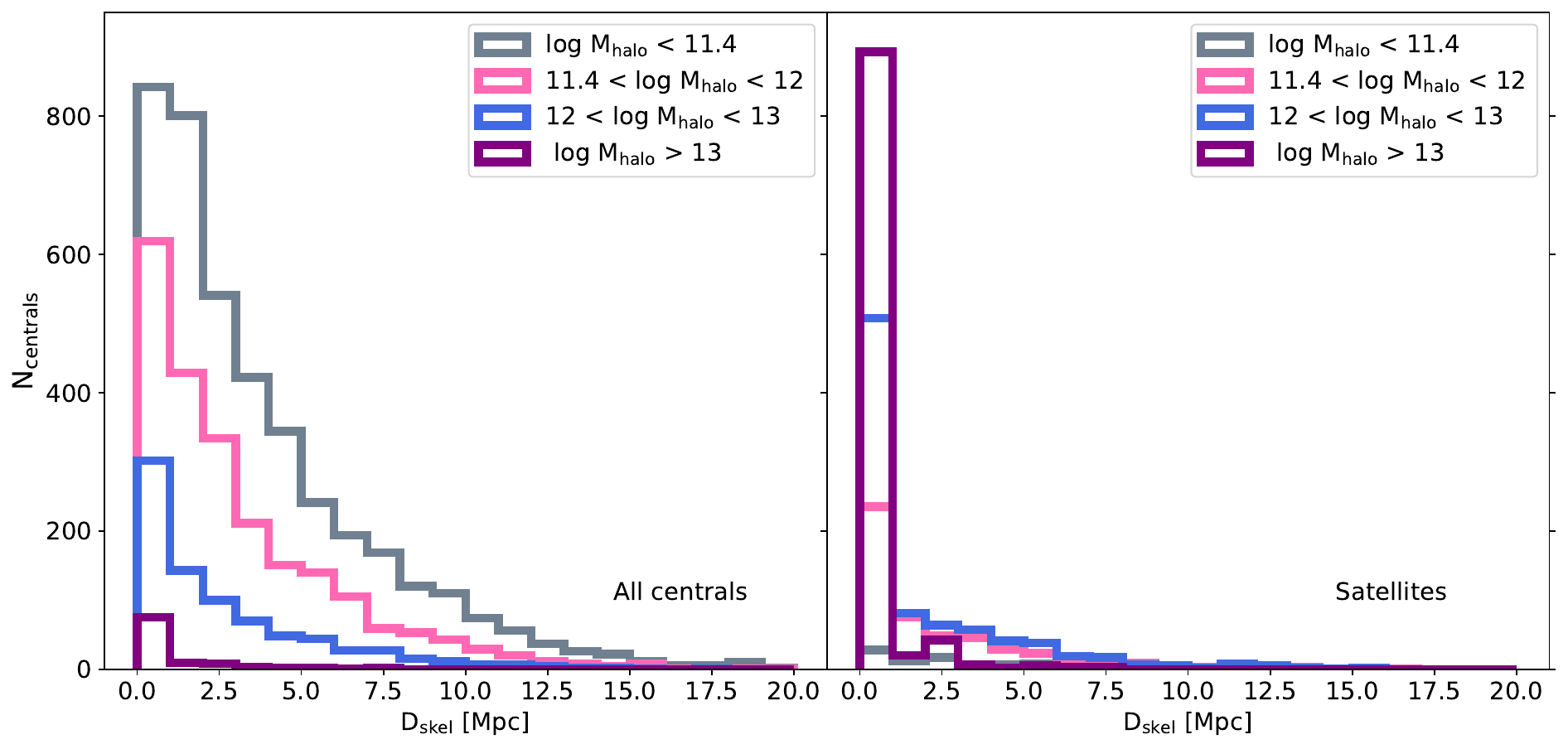}
\caption{Histograms showing the distribution of centrals (left) and satellites (right) in halo mass bins with distance to filament. Satellites are largely located very close to filaments in all halo mass bins.}
\label{fig:mhalohist}
\end{figure*}

Our results have shown that the increasing trends of stellar mass, colour/red-sequence fraction and gas-poor fraction closer to filaments are more significant for galaxies in groups than for single galaxies implying that the group environment effects dominate over effects due to the filament environment alone. However, an outstanding question is why we see trends with distance from filament in the group galaxy samples even after separation into stellar mass bins? Figure~\ref{fig:filaments} and Figure~\ref{fig:mhalohist} hint at the answer to this question. Figure~\ref{fig:filaments} shows the locations of central galaxies colour- and size-coded by the halo mass of their groups. The filament backbones follow the structures outlined by the highest mass haloes. Figure~\ref{fig:mhalohist} illustrates this more quantitatively, showing the distribution of central galaxies with distance to filament in bins of group halo mass. The vast majority of the highest mass haloes (log~(\Mhalo~/~\Msun)~$>13$) are found within 1 Mpc of the filaments. 

\begin{figure*}
\includegraphics[width=2\columnwidth]{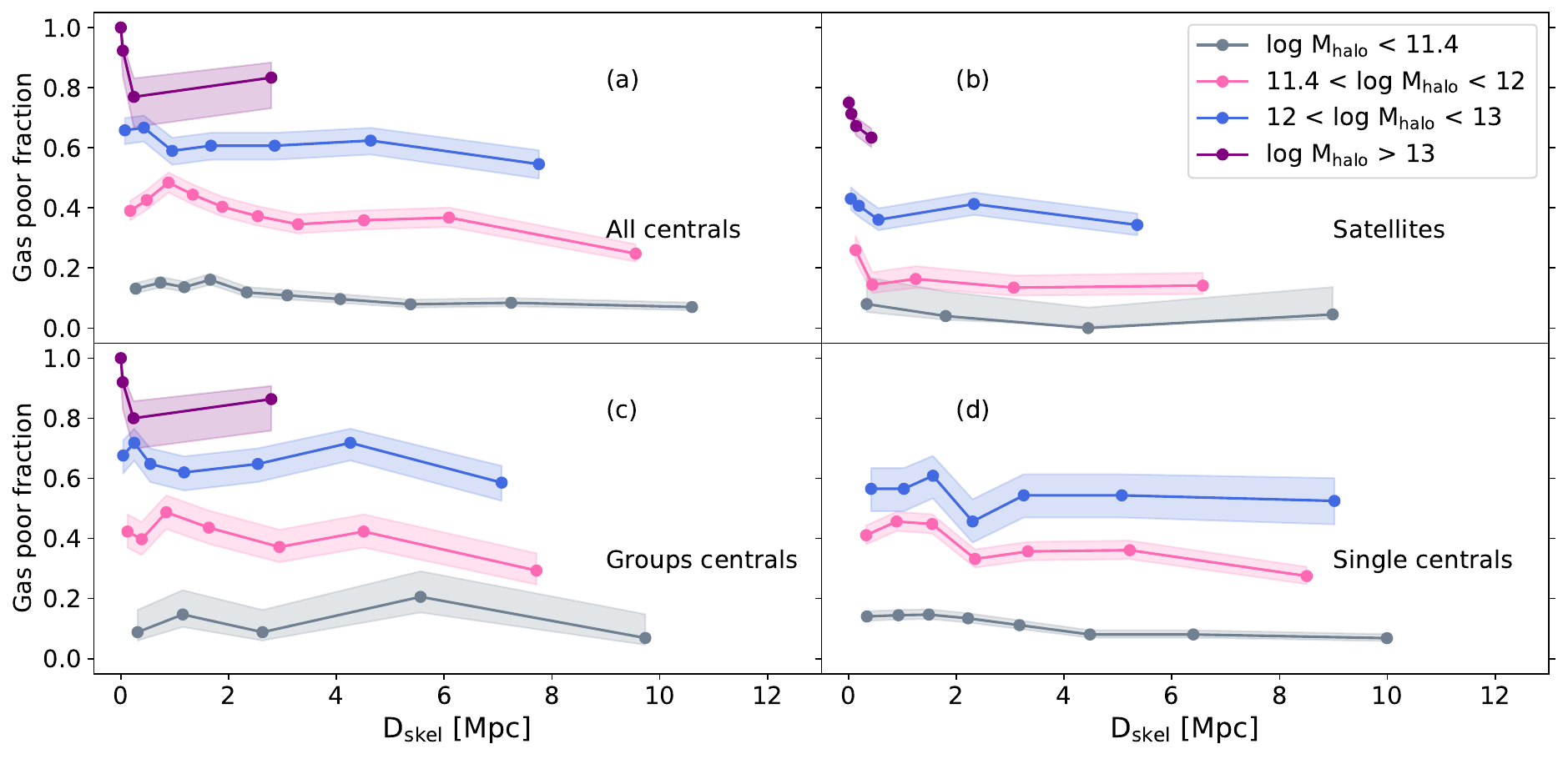}
\caption{The gas-poor fraction for central galaxies vs distance to filament in bins of halo mass. The gas-poor fraction for log (\Mhalo / \Msun) < 11.4 is shown in gray, 11.4< log (\Mhalo / \Msun) < 12 is shown in blue, 12 < log (\Mhalo / \Msun) < 13 is shown in pink and log (\Mhalo / \Msun) > 13 is shown in purple. Panel \textit{(a)} shows all central galaxies. Panel \textit{(b)} shows satellite galaxies. Panels \textit{(c)} and \textit{(d)} show central galaxies for groups \textit{(c)} and singles \textit{(d)}. There is a statistically significant increase in the gas poor fraction for single, central galaxies in low mass (\Mhalo /\Msun < 11.4) halos. There are no statistically significant trends with distance to filaments in any halo mass bins for group centrals.}
\label{fig:gaspoorcents}
\end{figure*}

In an effort to disentangle the contributions from groups, filaments and haloes on the galaxy properties, we present in Figure~\ref{fig:gaspoorcents}, the gas-poor fraction vs \Dskel\ for all central galaxies (panel a), group centrals (panel c), single centrals (panel d), and satellites (panel b) binned by halo mass. The binning in halo mass follows that used in \citet{stark2016resolve}. Although the \citet{stark2016resolve} halo masses correspond to our galaxy stellar mass regimes via the central galaxy \mstar-\Mhalo relation, there are satellites with lower stellar masses and scatter within the relation. We find a small increase in the gas-poor fraction of central galaxies in low mass haloes (panel a).

When separated into group and single central galaxies, we find no significant trends in any halo mass bin for group centrals (panel c). For single central galaxies (panel d), we find a small, marginally-significant trend for galaxies in the lowest halo mass bin. Galaxies in this halo mass bin span a range of stellar masses (approximately  8 < log (\mstar/\Msun) < 10) that is broader than the lowest stellar mass bin shown in earlier figures due to the inclusion of satellite and central galaxies which are scattered off the \mstar-\Mhalo relation. Given this range in masses, it is unsurprising that the trend seen here is consistent with the statistically significant trend seen in Figure \ref{fig:gaspoorodekoncomp} for the broader mass bin chosen to match \citet{odekon2018effect}’s selection. There are only two single galaxies with log~(\Mhalo~/~\Msun)~>~13, therefore this bin is not shown for single centrals. We find a statistically significant increase in gas-poor fraction for satellites in the highest halo mass bin (panel b). However, because all these satellites fall within 2.5 Mpc of the filaments, these satellites are likely all part of groups that fall within filaments and thus we cannot comment on the effect of filaments themselves for these galaxies. 

We note that satellite galaxies are not as gas-poor as central galaxies in the same mass haloes. However, this may be due to satellite galaxies on average having lower stellar masses than their central counterparts.

The gas-poor fraction for the satellites in the highest halo mass bin is much higher than in other halo mass bins and these galaxies are all located in haloes very close to filaments, i.e., with very low \Dskel\ values.  
Together, the satellite and group central galaxies make up the `group' galaxy sample and the single centrals correspond to the `single' galaxies studied earlier. Therefore, it must be these satellite galaxies in the highest mass haloes, which are only found very close to filaments, that are dominating the gas-poor fraction of group galaxies at low \Dskel\ values leading to the overall steep increase in the gas-poor fraction of group galaxies close to filaments seen in Figures \ref{fig:gaspoorfracgroupssingles} and~\ref{fig:gaspoorcomp} i.e., the increasing trend in gas-poor fraction close to filament seen for group galaxies is simply due to the location of the most massive haloes close to filaments. 

Another interpretation of the reduced gas content of galaxies in low mass haloes is that these haloes may have been subjected to `fly-by' interactions. In this scenario, smaller haloes fall into larger haloes, and are subjected to gas stripping before `splashing back' by leaving the larger haloes \citep{Gill2005splashback,McBride2009splashback}. \citet{stark2016resolve} identified this as a mechanism for low-mass haloes in overdense regions in their analysis of the RESOLVE survey. As filaments are considered overdense regions, it is possible that this mechanism is at play close to filaments. 

Overall, our results suggest that for single central galaxies in low-mass haloes, the denser filament environment may be a site of gas-removal processes in addition to halo-related processes. These results are aligned with recent works by \citet{Song2021filaments} and \citet{Winkel2021quenching} which examined the effects of filaments in addition to the effects of haloes. 
However, the effects of the filament environment within group galaxies, both centrals and satellites, seem negligible.

\subsubsection{The effect of nodes}
\label{sec:nodes}
As indicated in Section ~\ref{subsec:groupssingles}, we separate galaxies into groups and singles to isolate the effect of filaments from the influence of group and other high density environments. Previous works take the approach of removing nodes, which are regions of high density, identified by DisPerSE as critical points of type-3 in three dimensions, that may introduce different effects on galaxies due to their different structure, density, velocity fields and tidal forces on galaxies compared to filaments. Based on the Horizon-AGN simulations, a radius of 3.5 Mpc was used by \citet{Laigle2018cosmos} to remove galaxies close to nodes in 3-D studies of filaments. We examine the effect of removing node galaxies in addition to our removal of group galaxies. In practice, removing group galaxies already removes a large percentage of ‘node galaxies’ (2301 node galaxies are within groups), with only 744 single node galaxies which we also remove for this test. Plots corresponding to Figures \ref{fig:avemass}, \ref{fig:mstargroupssingles}, \ref{fig:redfraccomp} and \ref{fig:gaspoorcomp} with node galaxies removed can be found in Appendix \ref{appNodes}.

We find that removing node galaxies has a negligible effect on the previously measured trends in stellar mass; the median stellar mass is still higher close to filaments, as shown in Figure \ref{fig:avemassnodes}. Removing node galaxies decreases the trend in red fraction for intermediate mass galaxies such that it is no longer statistically significant. Similarly, because many group galaxies fall within nodes, the trend in red fraction for galaxies in groups, except for the highest mass bin, is no longer statistically significant when nodes are removed. We no longer find a statistically significant increase in the red fraction for single, low-mass galaxies close to filaments (see Figure \ref{fig:redfraccompnodes}). 

Removing node galaxies introduces a statistically significant increase in the gas-poor fraction for low mass, single galaxies, however, and removes the statistically significant trend for high mass single galaxies. This result - particularly for low mass galaxies - agrees with our overall conclusions that there may be evidence for cosmic web stripping of low mass galaxies. The gas-poor fraction at low \Dskel for galaxies in groups is lower when nodes are removed, reducing the significance of the trends. These changes to the trends for group galaxies are not unexpected, because the majority of galaxies close to nodes are group galaxies.

\section{Summary and conclusion} \label{sect:conclusion}

We have presented an analysis of the stellar mass, \textit{u-r} colour and gas properties of galaxies with respect to their distance to cosmic web filaments, group and halo environments in the ECO survey. In summary:

\begin{itemize}

    \item Galaxies, both singles and in groups, have higher stellar masses closer to filaments.
    \item The fraction of red galaxies is higher close to filaments.However, when galaxy groups and stellar mass are accounted for, we find that the red fraction is higher close to filaments regardless of their stellar mass and that only low mass single galaxies show a statistically significant increase in their red fraction due to filaments.
    \item Galaxies are more gas-poor close to filaments. Low and intermediate mass group galaxies are more gas-poor close to filaments. Low mass single galaxies show an increase in gas-poor fraction close to filaments. Single galaxies with 8.5 < log (\mstar/\Msun )< 10.5. also show an increase in the gas-poor fraction close to filaments, in agreement with previous works \citep{odekon2018effect}. 
    \item The increasing trends in stellar mass, red-fraction and gas-poor fraction closer to filaments seen for group galaxies are mainly driven by the fact that the highest mass haloes are preferentially located within or close to the filaments. This is supported by the fact that there are few observed trends of gas-poor fraction with distance from filament for group galaxies (either centrals or satellites) in similar halo mass bins. Therefore, their group environment seems to dominate over any filament effects on their evolutionary processes.
    \item Groups and nodes have a stronger effect on the reddening of galaxies than filaments, but filaments still have a small effect on the gas content of low mass, single galaxies. This may indicate possible cosmic web stripping. 

\end{itemize}

In this work, we have shown that although group and filament environment play a role in the evolutionary process, the influence of group environment is far more pronounced. One of the strengths of the ECO and RESOLVE datasets is the ability to cleanly separate out group environments to test the effects of filaments independently. These results are important for understanding the growth of galaxies in stellar mass as they travel along filaments to high density clusters and understanding the mechanisms which affect the accretion and stripping of gas from galaxies, leading to quenching.

Although a detailed comparison of the G3 and friends-of-friends group finders in terms of cosmic web filaments is beyond the scope of this work, comparing the effects of different group-finding techniques in the future may provide insight into the complex interplay between group, cluster and filament environments on galaxy properties.
Cutting-edge surveys measuring the gas content of galaxies at higher and higher redshifts, like those underway with the MeerKAT array (for example, LADUMA \citep{Blyth2016LADUMA} and MIGHTEE \citep{jarvis2017MIGHTEE}) will allow us to understand the evolution of the cosmic web across cosmic time. 

%\begin{equation}
 %   x=\frac{-b\pm\sqrt{b^2-4ac}}{2a}.%
%	\label{eq:quadratic}
%\end{equation}

%\begin{figure}
%	\includegraphics[width=\columnwidth]{example}
 %   \caption{This is an example figure. Captions appear below each figure.
%	Give enough detail for the reader to understand what they're looking at,
%	but leave detailed discussion to the main body of the text.}
 %   \label{fig:example_figure}
%\end{figure}

%\begin{table}
%	\centering
%	\caption{This is an example table. Captions appear above each table.
%	Remember to define the quantities, symbols and units used.}
%	\label{tab:example_table}
%	\begin{tabular}{lccr} % four columns, alignment for each
%		\hline
%		A & B & C & D\\
%		\hline
%		1 & 2 & 3 & 4\\
%		2 & 4 & 6 & 8\\
%		3 & 5 & 7 & 9\\
%		\hline
%	\end{tabular}
%\end{table}

%\section{Conclusions}

\section*{Acknowledgements}

MH would like to acknowledge support from the South African Radio Astronomical Observatory doctoral scholarship programme. SJK and ZLH acknowledge support from NSF award AST-1814486 and SJK and DSC acknowledge support from AST-2007351. ZLH is supported through a North Carolina Space Grant Graduate Research Fellowship. We acknowledge the use of the ilifu cloud computing facility – www.ilifu.ac.za, a partnership between the University of Cape Town, the University of the Western Cape, Stellenbosch University, Sol Plaatje University and the Cape Peninsula University of Technology. The ilifu facility is supported by contributions from the Inter-University Institute for Data Intensive Astronomy (IDIA – a partnership between the University of Cape Town, the University of Pretoria and the University of the Western Cape), the Computational Biology division at UCT and the Data Intensive Research Initiative of South Africa (DIRISA)

%%%%%%%%%%%%%%%%%%%%%%%%%%%%%%%%%%%%%%%%%%%%%%%%%%
\section*{Data Availability}

 The RESOLVE and ECO datasets used in this work is available in \citet{Hutchens2023G3}. Previous ECO and RESOLVE data releases can be found on the RESOLVE website at \url{https://resolve.astro.unc.edu/}.



\bibliographystyle{mnras}
%\bibliography{thesisbib} % if your bibtex file is called example.bib
\bibliography{ECOmain}

%\begin{thebibliography}{99}
%\end{thebibliography}

%%%%%%%%%%%%%%%%%%%%%%%%%%%%%%%%%%%%%%%%%%%%%%%%%%

%%%%%%%%%%%%%%%%% APPENDICES %%%%%%%%%%%%%%%%%%%%%

\appendix

\section{The Effect of Nodes}
\label{appNodes}

Figure showing the distribution of nodes, node galaxies and galaxies groups. 

\begin{figure*}
\centering
    \includegraphics[width=0.8\linewidth]{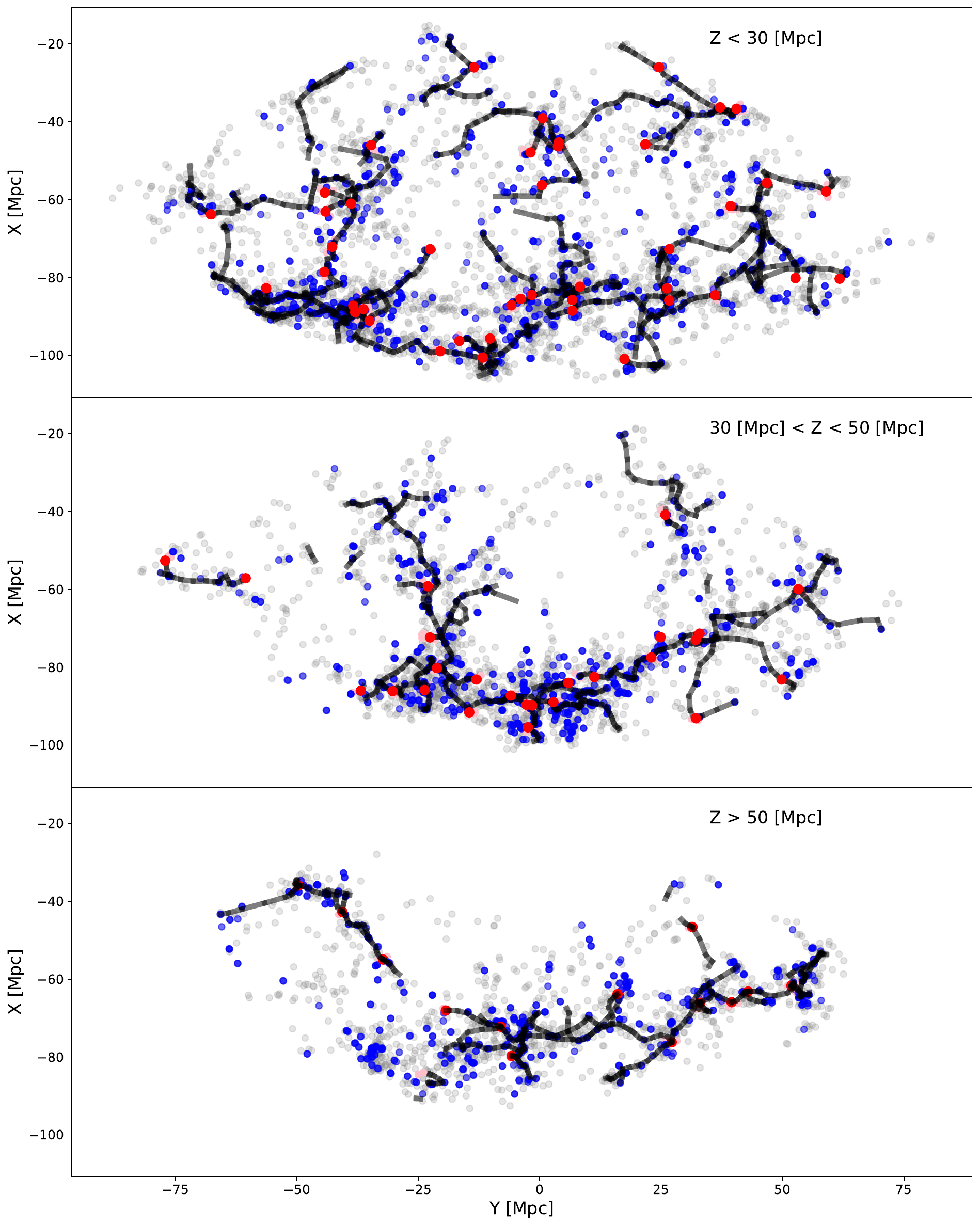}
\caption{Filaments in ECO presented in slices along the Z-axis. The top panel shows the slice where Z < 30 Mpc. The middle panel shows 30 Mpc < Z < 50 Mpc. The lower panel shows Z > 50 Mpc. Nodes are indicated by red circles, with node galaxies (i.e galaxies within 3.5 Mpc of nodes) indicated by pink dots. Galaxies which belong to groups are shown as blue circles. The remaining single galaxies are shown in gray. }
\label{fig:filamentsnodes}
\end{figure*}

Figures showing the trends in stellar mass, red fraction and gas-poor fraction when galaxies within 3.5 Mpc of nodes are removed. 

\begin{figure*}
\includegraphics[width=2\columnwidth]{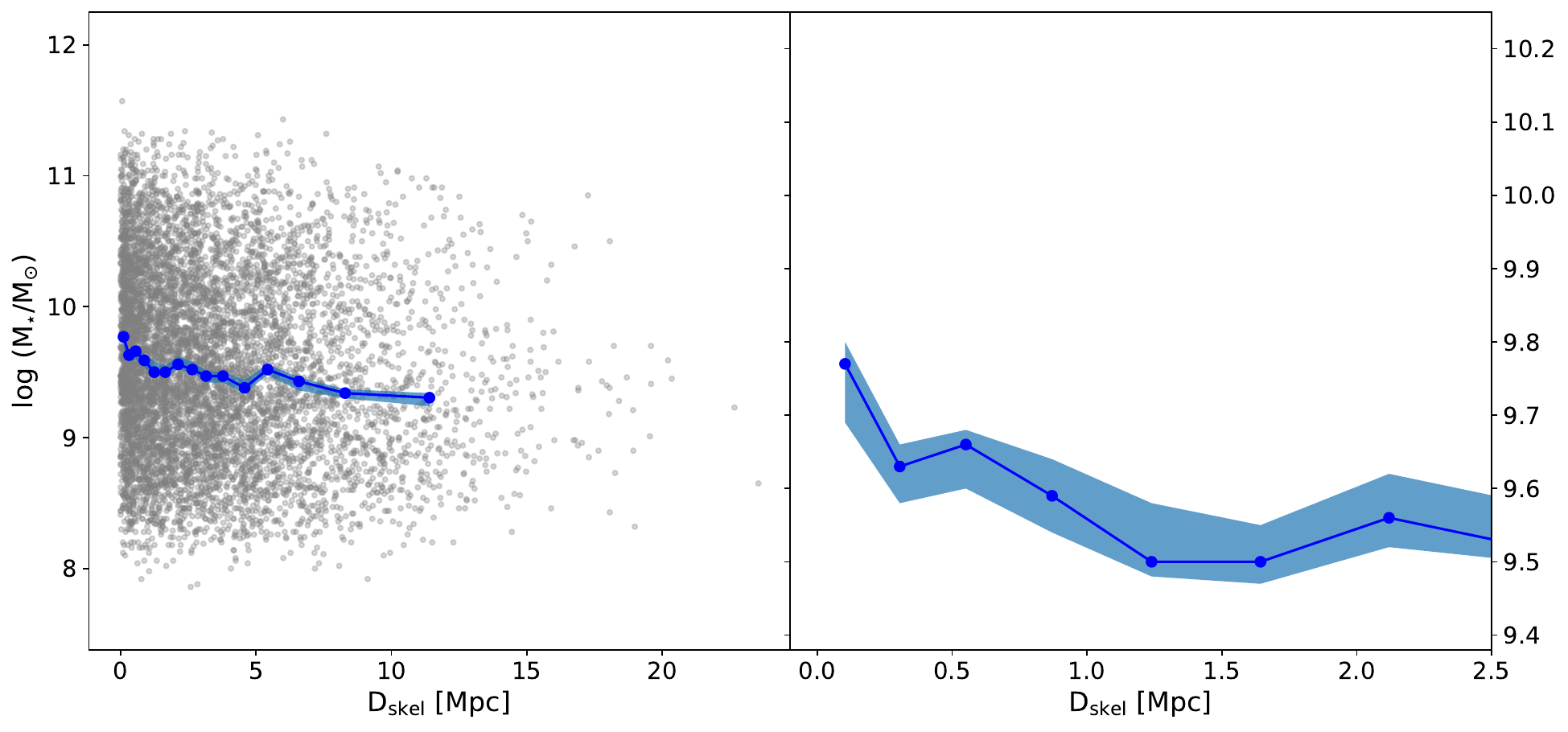}
\caption{Distribution of stellar masses as a function of distance to filament (\Dskel) when node galaxies are removed. Individual galaxies are shown in grey. The median stellar mass in each distance bin is shown in blue. The coloured band indicates the 1$\sigma$ error on the median. The right panel shows this distribution for 0 Mpc < \Dskel < 2.5 Mpc and is zoomed-in on the y-axis to highlight the behaviour close to the filaments. The median stellar mass decreases as distance to filament increases.}
\label{fig:avemassnodes}
\end{figure*}

\begin{figure*}
\includegraphics[width=2\columnwidth]{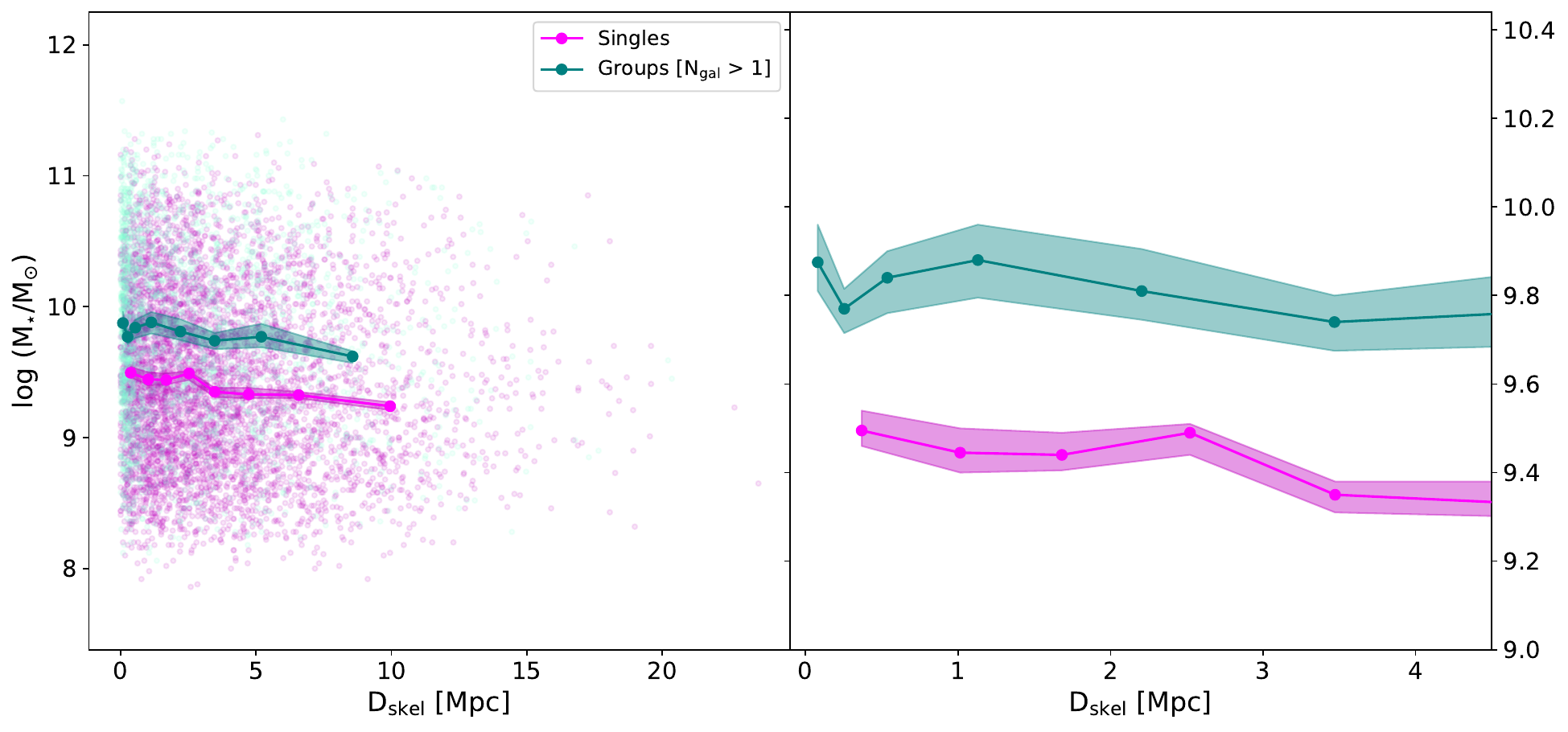}
\caption{The median stellar mass vs distance to filament (\Dskel) for galaxies in groups (teal) and single galaxies (magenta) when node galaxies are removed.  Galaxies in groups have systematically higher stellar masses than single galaxies. There is a small decrease in the median stellar mass for both group and single galaxies as distance to filament increases. The right panel shows this distribution for 0 Mpc < \Dskel < 4 Mpc and the 9.0 < log (\mstar/\Msun) < 10.4 range on the y-axis to highlight the behaviour close to the filaments.}
\label{fig:mstargroupssinglesnodes}
\end{figure*}

\begin{figure*}
\includegraphics[width=2\columnwidth]{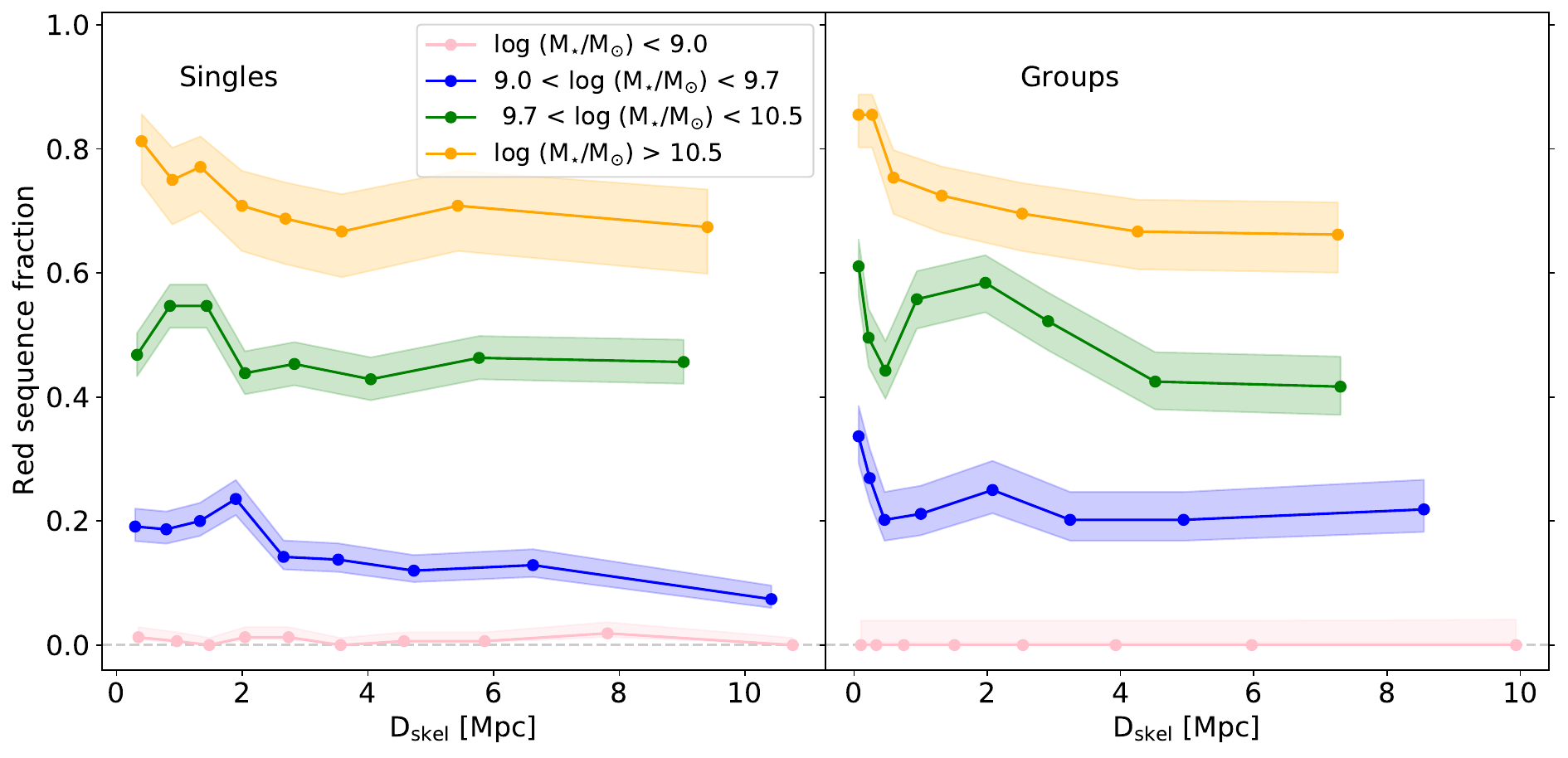}
\caption{The red fraction vs distance to filament is shown for single galaxies (left panel) and group galaxies (right panel) for low, intermediate and high stellar mass sub-samples when nodes are removed. Group galaxies show a statistically significant increase in red fraction with decreasing \Dskel for high mass galaxies. There are no statistically significant trends in red fraction with distance to filaments for single galaxies in any mass bin. }
\label{fig:redfraccompnodes}
\end{figure*}

\begin{figure*}
\includegraphics[width=2\columnwidth]{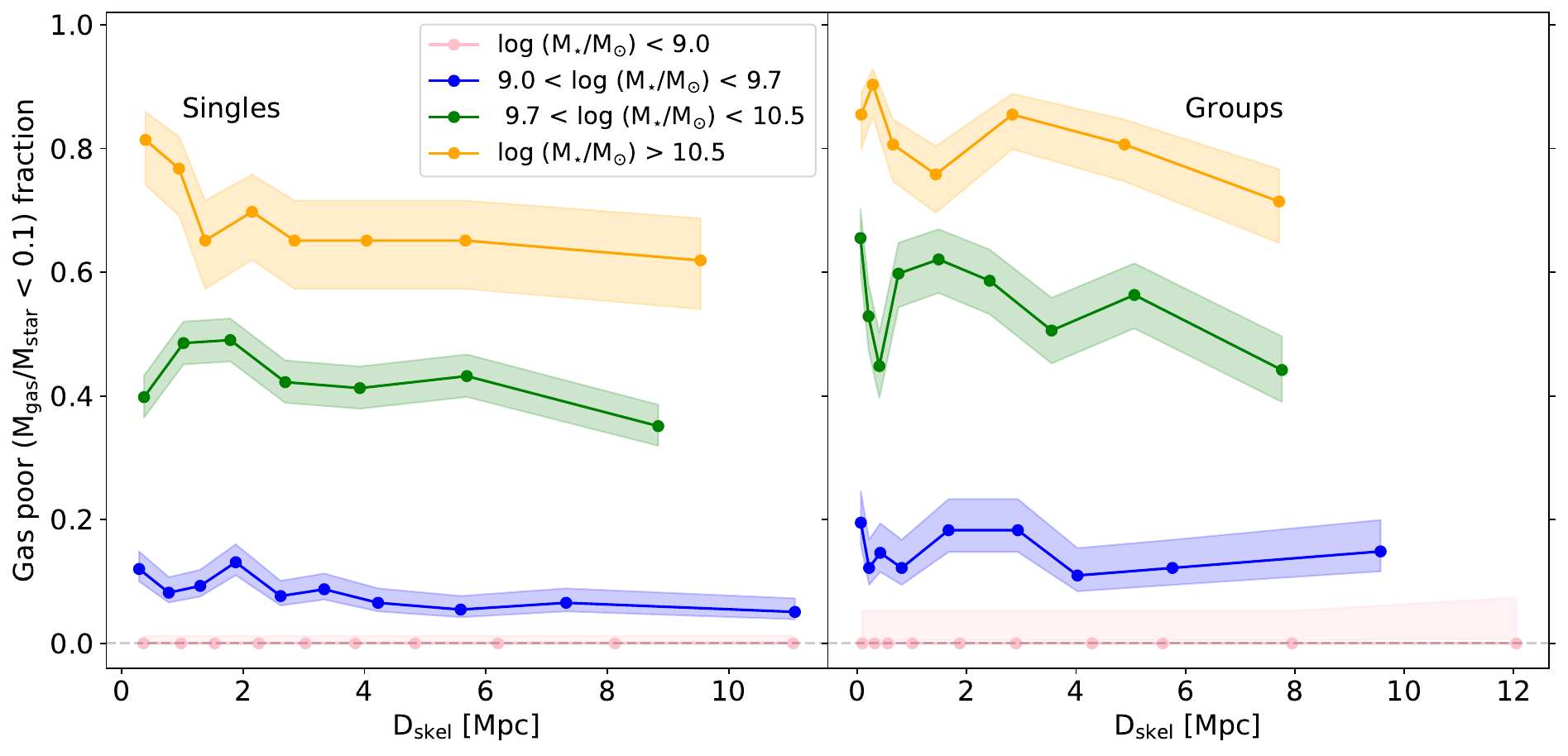}
\caption{Gas-poor fraction vs distance to filament for single galaxies (left panel) and group galaxies (right panel) for low, intermediate and high stellar mass sub-samples when node galaxies are removed. None of the trends are found to be statistically significant for galaxies in groups. Low mass single galaxies show statistically significant increases in gas-poor fraction with decreasing \Dskel. }
\label{fig:gaspoorcompnodes}
\end{figure*}

\section{Statistical Significance}
\label{appB}

Tables showing the correlation co-efficient and p-values calculated for trends in Section \ref{sect:results} and \ref{sect:discussion}. These values are calculated using the Spearman's Rank Test. Trends are considered statistically significant if p < 0.003.

\begin{table*}
\caption{Table showing the correlation coefficient and p-values calculated for trends in Section \ref{sect:results}. The columns indicate the name of the figure, the figure number, stellar mass or group/single bin, correlation coefficients and p-values. }
\label{tab:spearmans}

%\resizebox{\textwidth}{!}{%
%\begin{tabular}{lllll}
\begin{tabular}{lclll}
\textbf{Plot }                               & \textbf{Figure Number   }                  & \textbf{Mass Bin }    & \textbf{Spearman's $\rho$} &\textbf{ p value}  \\\hline \hline
Red sequence fraction               & \ref{fig:redfrac}                 & Ultra-dwarf  & -0.23   & 5.30E-01 \\
                                    &                                   & Low          & -0.88   & 8.14E-04 \\
                                    &                                   & Intermediate & -0.92   & 1.80E-04 \\
                                    &                                   & High         & -0.98   & 2.29E-07 \\
Gas-poor fraction                   & \ref{fig:gaspoorfrac}             & Ultra-dwarf  & 0.00    & 0.00E+00 \\
                                    &                                   & Low          & -0.88   & 8.14E-04 \\
                                    &                                   & Intermediate & -0.92   & 2.04E-04 \\
                                    &                                   & High         & -0.97   & 2.37E-06 \\
Red sequence fraction in groups and single & \ref{fig:redfracgroupssingles}         & Groups      & -0.95 & 2.93E-05 \\
                                    &                                   & Singles      & -0.90   & 3.44E-04 \\
Red sequence fraction - groups only        & \ref{fig:redfraccomp}              & Ultra-dwarf & -0.96 & 7.32E-06 \\
                                    &                                   & Low          & -0.90   & 3.44E-04 \\
                                    &                                   & Intermediate & -0.52   & 1.22E-01 \\
                                    &                                   & High         & -0.87   & 1.17E-03 \\
Red sequence fraction - singles only       & \ref{fig:redfraccomp}              & Ultra-dwarf & -0.89 & 6.81E-03 \\
                                    &                                   & Low          & -0.96   & 4.54E-04 \\
                                    &                                   & Intermediate & -0.12   & 7.42E-01 \\
                                    &                                   & High         & -0.85   & 8.07E-04 \\
Gas-poor fraction in groups and singles    & \ref{fig:gaspoorfracgroupssingles} & Groups      & -0.74 & 3.66E-02 \\
                                    &                                   & Singles      & -0.71   & 4.81E-02 \\
Gas-poor fraction - groups only     & \ref{fig:gaspoorcomp}             & Ultra-dwarf  & 0.00    & 0.00E+00 \\
                                    &                                   & Low          & -0.85   & 4.12E-03 \\
                                    &                                   & Intermediate & -0.90   & 2.01E-03 \\
                                    &                                   & High         & -0.94   & 1.85E-03 \\
Gas-poor fraction - singles only    & \ref{fig:gaspoorcomp}             & Ultra-dwarf  & 0.00    & 0.00E+00 \\
                                    &                                   & Low          & -0.80   & 5.21E-03 \\
                                    &                                   & Intermediate & -0.61   & 1.48E-01 \\
                                    &                                   & High         & -0.82   & 2.34E-02 \\
\textbf{Nodes }                              &                                   &              &         &          \\ \hline \hline
Red sequence fraction - groups only & \ref{fig:avemassnodes}            & Ultra-dwarf  & -       & -        \\
                                    &                                   & Low          & -0.54   & 1.70E-01 \\
                                    &                                   & Intermediate & -0.62   & 1.02E-01 \\
                                    &                                   & High         & -0.99   & 1.46E-05 \\
Red sequence fraction - singles only       & \ref{fig:avemassnodes}             & Ultra-dwarf & -0.12 & 7.42E-01 \\
                                    &                                   & Low          & -0.83   & 5.27E-03 \\
                                    &                                   & Intermediate & -0.54   & 1.68E-01 \\
                                    &                                   & High         & -0.83   & 1.14E-02 \\
Gas-poor fraction - groups only     & \ref{fig:mstargroupssinglesnodes} & Ultra-dwarf  & -       &          \\
                                    &                                   & Low          & -0.26   & 4.93E-01 \\
                                    &                                   & Intermediate & -0.47   & 2.05E-01 \\
                                    &                                   & High         & -0.67   & 9.76E-02 \\
Gas-poor fraction - singles only           & \ref{fig:mstargroupssinglesnodes}  & Ultra-dwarf & -     &          \\
                                    &                                   & Low          & -0.83   & 2.78E-03 \\
                                    &                                   & Intermediate & -0.32   & 4.82E-01 \\
                                    &                                   & High         & -0.88   & 4.42E-03

\end{tabular}%
%}
\end{table*}

\begin{table*}
\caption{Significance p-values and correlation coefficients (Spearman's $\rho$) for Figure \ref{fig:gaspoorcents} for each halo mass bin and panel.}
\label{tab:figure17}
\begin{tabular}{llll}
\textbf{Plot}           & \textbf{Halo Mass }                                           & \textbf{Spearman's $\rho$} & \textbf{P value } \\ \hline \hline

All centrals    & log M$_{\mathrm{halo}}$/\Msun \textless 11.4                  & -0.88            & 8.14E-04                    \\
                & 11.4 \textless log M$_{\mathrm{halo}}$/\Msun \textless 12     & -0.79            & 6.10E-03                    \\
                & 12 \textless log M$_{\mathrm{halo}}$ \Msun\textless 13        & -0.63            & 1.29E-01                    \\
                & log M$_{\mathrm{halo}}$ / \textbackslash{}Msun\textgreater 13 & -0.80            & 2.00E-01                    \\
Satellites      & log M$_{\mathrm{halo}}$/\Msun \textless 11.4                  & -0.40            & 6.00E-01                    \\
                & 11.4 \textless log M$_{\mathrm{halo}}$/\Msun \textless 12     & -0.80            & 1.04E-01                    \\
                & 12 \textless log M$_{\mathrm{halo}}$ \Msun \textless 13       & -0.70            & 1.88E-01                    \\
                & log M$_{\mathrm{halo}}$ / \Msun\textgreater 13                & -1.00            & 0.00E+00                    \\
Group centrals  & log M$_{\mathrm{halo}}$/\Msun \textless 11.4                  & -0.21            & 7.41E-01                    \\
                & 11.4 \textless log M$_{\mathrm{halo}}$/\Msun \textless 12     & -0.45            & 3.10E-01                    \\
                & 12 \textless log M$_{\mathrm{halo}}$ \Msun \textless 13       & -0.44            & 3.28E-01                    \\
                & log M$_{\mathrm{halo}}$ / \Msun\textgreater 13                & -0.80            & 2.00E-01                    \\
Single centrals & log M$_{\mathrm{halo}}$/\Msun \textless 11.4                  & -0.90            & 2.44E-03                    \\
                & 11.4 \textless log M$_{\mathrm{halo}}$/\Msun \textless 12     & -0.75            & 5.22E-02                    \\
                & 12 \textless log M$_{\mathrm{halo}}$ \Msun \textless 13       & -0.65            & 1.11E-01                   
\end{tabular}%

\end{table*}

\bsp	% typesetting comment
\label{lastpage}
\end{document}